\documentclass[twocolumn]{aastex62}

\usepackage{graphicx}	
\usepackage{amsmath}	
\usepackage{amsfonts}
\usepackage{amssymb}	
\usepackage{soul}
\usepackage{lineno}
\usepackage{tikz}
\usepackage{hyperref}
\usepackage{algorithm,algpseudocode}

\usetikzlibrary{shapes.geometric, arrows}
\tikzstyle{startstop} = [rectangle, rounded corners, minimum width=3cm, minimum height=1cm,text centered, draw=black]
\tikzstyle{io} = [trapezium, trapezium left angle=70, trapezium right angle=110, minimum width=3cm, minimum height=1cm, text centered, draw=black]
\tikzstyle{process} = [rectangle, minimum width=3cm, minimum height=1cm, text centered, draw=black]
\tikzstyle{decision} = [diamond, minimum width=3cm, minimum height=1cm, text centered, draw=black]
\tikzstyle{arrow} = [thick,->,>=stealth]

\usepackage{array}
\newenvironment{conditions}
  {\par\vspace{\abovedisplayskip}\noindent\begin{tabular}{>{$}l<{$} @{${}={}$} l}}
  {\end{tabular}\par\vspace{\belowdisplayskip}}

\submitjournal{ApJ}

\shorttitle{An unsupervised ML based algorithm for detecting and characterising WINQES}
\shortauthors{Bawaji et al.}

\begin{document}

\title{An unsupervised machine learning based algorithm for detecting Weak Impulsive Narrowband Quiet Sun Emissions and characterizing their morphology} 


\correspondingauthor{Surajit Mondal} 
\email{surajit.mondal@njit.edu}

\author{Shabbir Bawaji}
\affil{e4r, ThoughtWorks India}

\author{Ujjaini Alam}
\affil{e4r, ThoughtWorks India}

\author[0000-0002-2325-5298]{Surajit Mondal}
\affiliation{Center for Solar-Terrestrial Research, New Jersey Institute of Technology, 323 M L King Jr Boulevard, Newark, NJ 07102-1982, USA}

\author[0000-0002-4768-9058]{Divya Oberoi}
\affiliation{National Centre for Radio Astrophysics, Tata Institute of Fundamental Research, S. P. Pune University, Pune 411007, India}

\author{Ayan Biswas}
\affiliation{National Centre for Radio Astrophysics, Tata Institute of Fundamental Research, S. P. Pune University, Pune 411007, India}
\affiliation{Department of Physics, Engineering Physics \& Astronomy, Queen’s University, Kingston, Ontario K7L 3N6, Canada}
\affiliation{Department of Physics, Royal Military College of Canada, Kingston, Ontario K7K 7B4, Canada}



\begin{abstract}

The solar corona is extremely dynamic. 
Every leap in observational capabilities has been accompanied by unexpected revelations of complex dynamic processes. 
The ever more sensitive instruments now allow us to probe events with increasingly weaker energetics.
A recent leap in the low-frequency radio solar imaging ability has led to the discovery of a new class of emissions, namely Weak Impulsive Narrowband Quiet Sun Emissions \citep[WINQSEs;][]{mondal2020}.
They are hypothesized to be the radio signatures of coronal nanoflares and could potentially have a bearing on the long standing coronal heating problem.
In view of the significance of this discovery, this work has been followed up by multiple independent studies.
These include detecting WINQSEs in multiple datasets, using independent detection techniques and software pipelines, and looking for their counterparts at other wavelengths. 
This work focuses on investigating morphological properties of WINQSEs and also improves upon 
the methodology used for detecting WINQSEs in earlier works.
We present a machine learning based algorithm to detect WINQSEs, classify them based on their morphology and model the isolated ones using 2D Gaussians.
We subject multiple datasets to this algorithm to test its veracity.
Interestingly, despite the expectations of their arising from intrinsically compact sources, WINQSEs 
tend to be resolved in our observations.
We propose that this angular broadening arises due to coronal scattering. WINQSEs can, hence, provide  ubiquitous and ever-present diagnostic of coronal scattering (and, in turn, coronal turbulence) in the quiet sun regions, which has not been possible till date. 


\end{abstract}


\section{Introduction}
\label{intro}

The solar corona is well known to be extremely dynamic.
With each leap in the capabilities of available instrumentation, the Sun has revealed a whole new class of dynamics. 
For instance, recently, with the Solar Orbiter \citep{forveille2020}, it has now become possible to observe in exquisite detail the dynamics of small scale transients referred to as ``campfires" \citep{berghmans2021}.
Similarly the Parker Solar Probe \citep{fox2016} has revealed the ubiquitous presence of switchbacks in the solar wind \citep{kasper2019}.
The low frequency radio observations are no different. The advent of new generation radio instruments is leading to multiple discoveries of previously unappreciated aspects \citep{solar_review}. 
In this paper, we focus on studies of the weak impulsive narrowband quiet sun emissions \citep[WINQSEs;][]{mondal2020, mondal2021,rohit2022,mondal2023}.
Recently discovered using the Murchison Widefield Array, an SKA precursor, WINQSEs have been found to be ubiquitous in the quiet Sun regions at the metre wavelength band.

Reporting their discovery, \citet{mondal2020}, henceforth referred to as M20, hypothesised that the detected impulsive emissions are weaker cousins of the type III and/or type I radio bursts \citep{reid2014,sherry2021} and are likely related to the nanoflares \citep{parker1988}. 
Like earlier works, the hypothesis is that the nanoflares would accelerate electrons to energies higher than the thermal background, which would then emit plasma emission through various wave-particle and wave-wave interactions \citep{aschwanden2005}. 
Since the radio emission arises from a coherent emission process, it should be easier to detect the radio signatures as compared to the thermal signatures usually observed in the extreme ultraviolet and soft X-ray bands.
While this has been known for a while and several authors have searched for nanoflares through these radio signatures, earlier works were only limited to the active regions \citep[e.g.][etc.]{mercier1997,ramesh2013}. 
This is primarily because the imaging fidelity and dynamic range required to detect these emission from the quiet Sun was simply not available till recently.
With the availability of the data from the MWA and automated imaging tools like the Automated Imaging Routine for Compact Arrays for Radio Sun \citep[AIRCARS;][]{mondal2019}, the state-of-the-art in solar spectroscopic snapshot imaging at low radio frequencies has improved by multiple orders of magnitude.
This enabled M20 to detect these emissions for the first time.

However M20 used a rather simplistic technique to detect WINQSEs. 
In addition to its inherent shortcomings, this technique also limits further studies of WINQSEs. 
M20 extracted the flux density time series from fixed regions of size similar to the instrumental resolution chosen arbitrarily to tile the Sun.
They used this time series to identify WINQSEs. 
As the regions chosen do not seamlessly tile the Sun, the WINQSEs lying in the intervening regions are likely to be missed. 
These regions are defined arbitrarily and WINQSEs with smaller overlaps with these regions will contribute a lower flux density leading to an underestimate of the flux density of WINQSEs which are detected.
There will also be a population of WINQSEs which could have been detected if the region was centered on the feature, but were missed because the limited overlap between the two implied that the integrated flux density of the region did not exceed the detection threshold.
Though \citet{mondal2023}(henceforth refereed as M23) improved some aspects of the data analysis procedures used in M20, this issue afflicts their analysis as well.
Additionally this technique implicitly assumed WINQSEs to be compact and essentially unresolved, without first demonstrating it.
In fact, the size and morphology of WINQSEs are key parameters of interest for understanding their relationships to nanoflares and perhaps also the propagation effects in the inhomogeneous turbulent corona.
Studies of propagation effects have been limited to strong active emissions, but given that WINQSEs are ubiqutous on the quiet Sun and every present, they can potentially serve as a convenient tool for characterising scattering simultaneously over the entire corona.

The present work overcomes the limitations of M20 and M23 just pointed out and presents the first characterization of the observed morphology of WINQSEs.
We have designed, implemented and successfully used a robust { unsupervised machine learning} based pipeline on multiple data-sets to identify WINQSEs and characterise their morphology using a generic 2D Gaussian model.
This work is organised as follows: 
Section \ref{sec:obs} describes the two datasets used here along with the pre-processing steps involved in preparing the radio images for the machine learning based pipeline, which is described in detail in Sec. \ref{sec:aiml}.
The results from the pipeline are presented in Sec. \ref{sec:results} followed by their discussion in Sec. \ref{sec:disc} and finally the conclusions in Sec. \ref{sec:conclusion}.

\section{Observations and Radio Imaging}
\label{sec:obs}

For this project, we have analysed MWA data from November 27, 2017 and June 20, 2020 obtained under the proposal code G0002. We refer to these datasets as 20171127 and 20200620 respectively. The following subsections discuss the state of the Sun on these days along with some details of the datasets and the analysis done. 


\subsection{Dataset 20171127}
\label{subsec:DS1}

These data are described in detail in M20.
Here we only present some key aspects of these observations and data analysis.

These data were taken using the Murchison Widefield Array Phase II extended configuration \citep{tingay2013,wayth2018} from 1:30--3:38 UT. This observation duration was broken into several chunks, each of duration four minutes.
There was only one active region (NOAAA 12689) on the visible part of the solar disc on this day. 
A single GOES B2.9 class X-ray flare was reported in the SWPC event list on this day, which took place outside our observing window. 
There was no evidence of any radio bursts in the data from the Culgoora spectrograph operating between 18--1800 MHz and the Learmonth spectrograph operating from 25--180 MHz.

\begin{figure*}
\centering
    \includegraphics[clip,trim={2cm 0 2.5cm 1cm},scale=0.35]{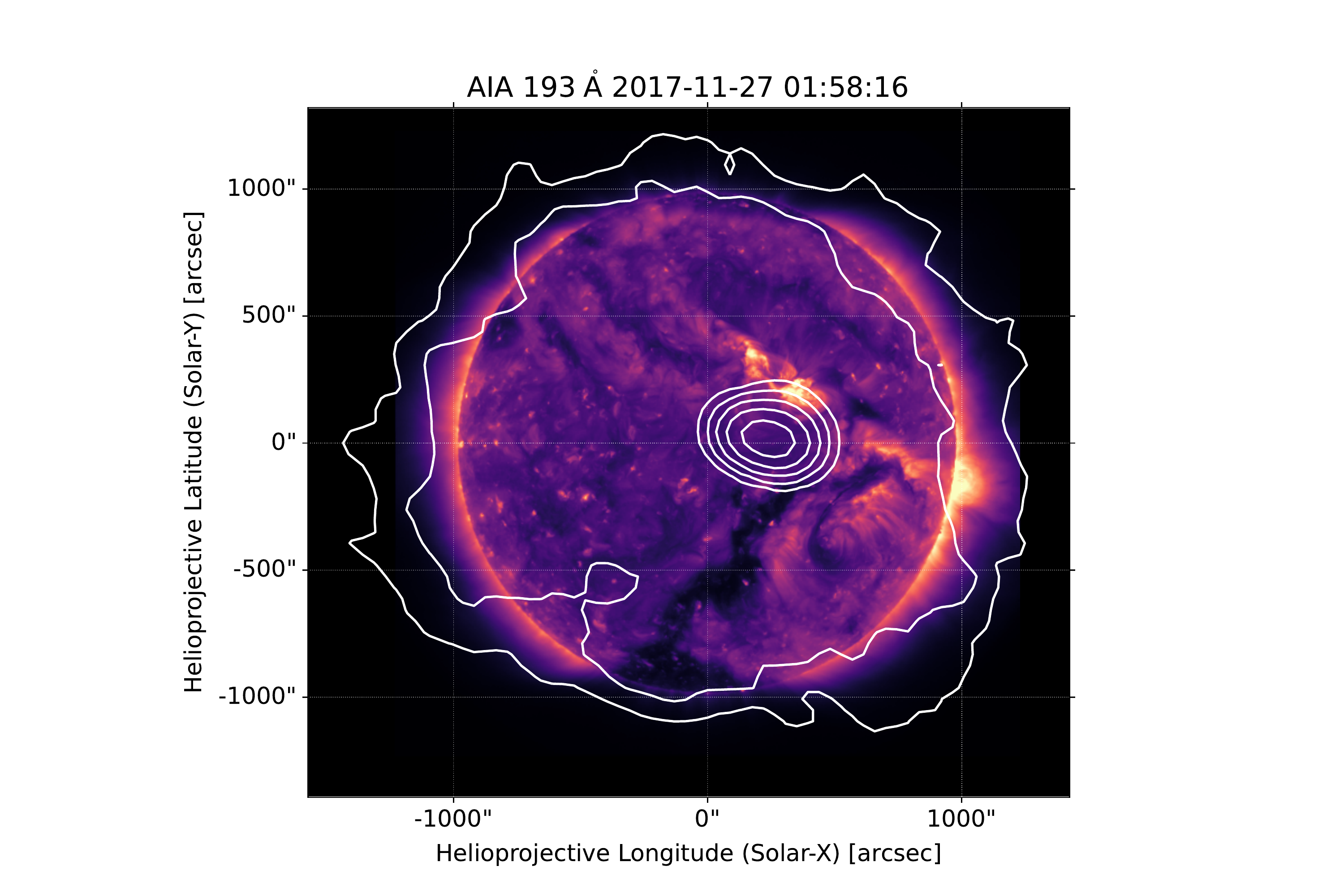}
    \includegraphics[clip,trim={4cm 0 2.5cm 1cm},scale=0.35]{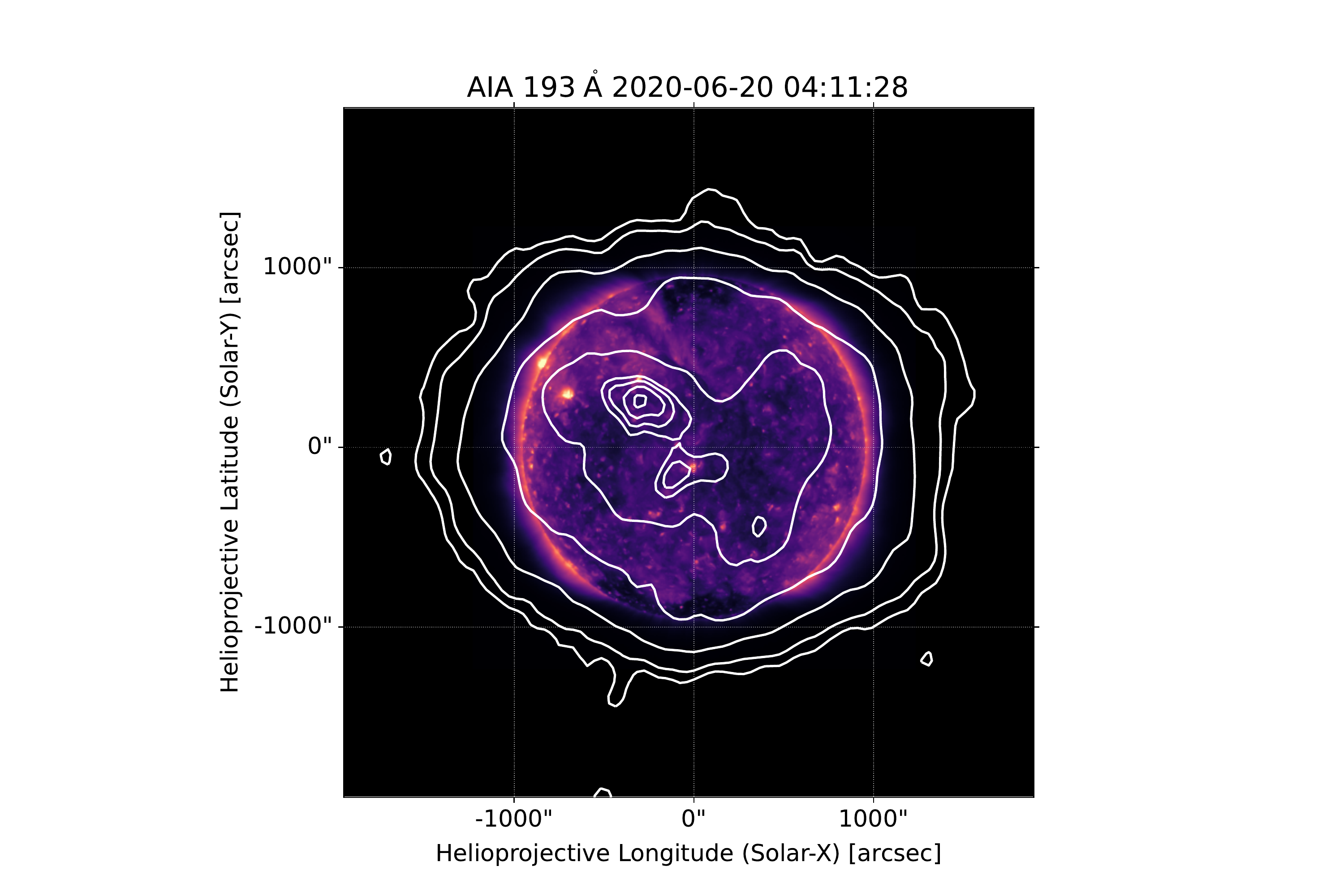}
    \caption{Left panel: An example 132 MHz radio image is shown in contours, overlaid on an AIA 193 \AA $\,$ image taken near the middle of the observing duration on 2017/11/27. 
    The contour levels are 0.006, 0.01, 0.04, 0.08, 0.16, 0.32, 0.64 times the peak in the radio image. The bright compact emission comes the type I noise storm associated with the only active region present on the visible solar disc.
    Right panel: The contours show an example 133 MHz radio image, overlaid on an AIA 193 \AA $\,$ image taken near the middle of the observation on 2020/06/20. The contours are at 0.04, 0.08, 0.2, 0.4, 0.6, 0.72, 0.74, 0.76, 0.78, 0.8, 0.9 times the peak in the radio image. }
    \label{fig:solar_image}
\end{figure*}

M20 analysed 70 minutes of data at four frequencies near 96, 120, 132, and 160 MHz using the Automated Imaging Routine for Compact Array for Radio Sun \citep[AIRCARS;][]{mondal2019}. The images were made every 0.5 s with a temporal resolution of 0.5 s and spectral resolution of 160 kHz. 
Here we focus on images at 132 MHz, the spatial resolution for which ranged between $\sim 2.5^{'}$--$4.2^{'}$. 
The typical dynamic range for these images is $\sim$800, and the most prominent feature is the type I noise storm associated with the lone active region on the visible solar disc.
The large angular scale quiet sun disc is detected with a signal to noise ratio of $\gtrsim 7$ except for a few time slices when the type I emission was very bright. These images are excluded from further analysis. An example image shown in the left panel of Fig. \ref{fig:solar_image} demonstrates the high fidelity of these images.

\subsection{Dataset 20200620}
\label{subsec:DS2}

These data and details of the analysis performed are described in detail in M23 and only some key aspects of these observations and data analysis are presented here.

These data were also taken using the extended configuration of the Murchison Widefield Array Phase II 
from 3:35--4:44 UT. Unlike 20171127, these data provided a contiguous sampling of the band from $\sim$119--151 MHz. 
The observation duration was broken into five minute chunks. 
The sun was very quiet on this day. There was no active region on the visible part of the solar disc and no events were reported in the SWPC event list. 
There was also no evidence of any radio bursts in the data from the Culgoora and the Learmonth spectrographs. 

We have analysed these data at four frequencies centered at 120.52, 128.20, 135.90 and 143.60 MHz. The images were made every 0.5 s with a temporal and spectral resolution of 0.5 s and 160 kHz respectively. 
The spatial resolution of the images was $\sim 3^{'}\times4^{'}$, with minor image-to-image variations. The quiet sun was detected without exception in all of the time slices with a signal to noise ratio of $\gtrsim 12$. 
An example high fidelity image from these data is shown in the right panel of Fig. \ref{fig:solar_image}.


\subsection{Generating datacubes for machine learning processing} \label{sec:gen_datacube}

While the fidelity of individual AIRCARS images is quite sufficient for this study, a few additional aspects need to be taken into account to turn the entire set of images into suitable 3D datacubes.
These issues were realized subsequent to the discovery of WINQSEs reported in M20, during the course of our continued investigations of WINQSEs and efforts to understand the low level effects seen in the data.
The aspects needed to be accounted for are:
\begin{enumerate}
    \item The raw images from AIRCARS can show low level jumps in the apparent brightness levels between adjacent data chunks.
    \item Small variations in spatial resolutions can arise between images as sometimes a few of the antennas calibrated with insufficient signal-to-noise drop out during imaging. These antennas tend to be in the outer sparse parts of the array and hence tend to have a larger impact on the size of the point spread function (PSF).
    These variations in spatial resolution lead to corresponding variations in flux density per beam, in turn giving rise to variations in pixel based values in the image.
    This issue usually arises only when imaging a very quiet featureless Sun which is the case for this dataset. 
\end{enumerate}
 
These variations lie in the range of 10--15\% and are usually not important for most solar radio studies.
However, for studies of the intrinsically weak WINQSEs, it becomes important to correct for them. 
It was found that matching the median apparent integrated solar flux density across the 4--5 minute data chunks, using a multiplicative factor, was sufficient to remove the low level jumps at data boundaries.
This ensures that all of the images are on the same flux density scale.
In order to match the resolution across different images, all of the images were smoothed to a common resolution using the CASA \citep{casa} task {\em imsmooth}.
The final spatial resolution for 20171127 and 20200620 datasets correspond to circles of size $250^{''}$ and $280^{''}$ respectively.

\section{Image Characterization Pipeline}
\label{sec:aiml}

\begin{figure}
\centering
    \includegraphics[width=7.5cm]{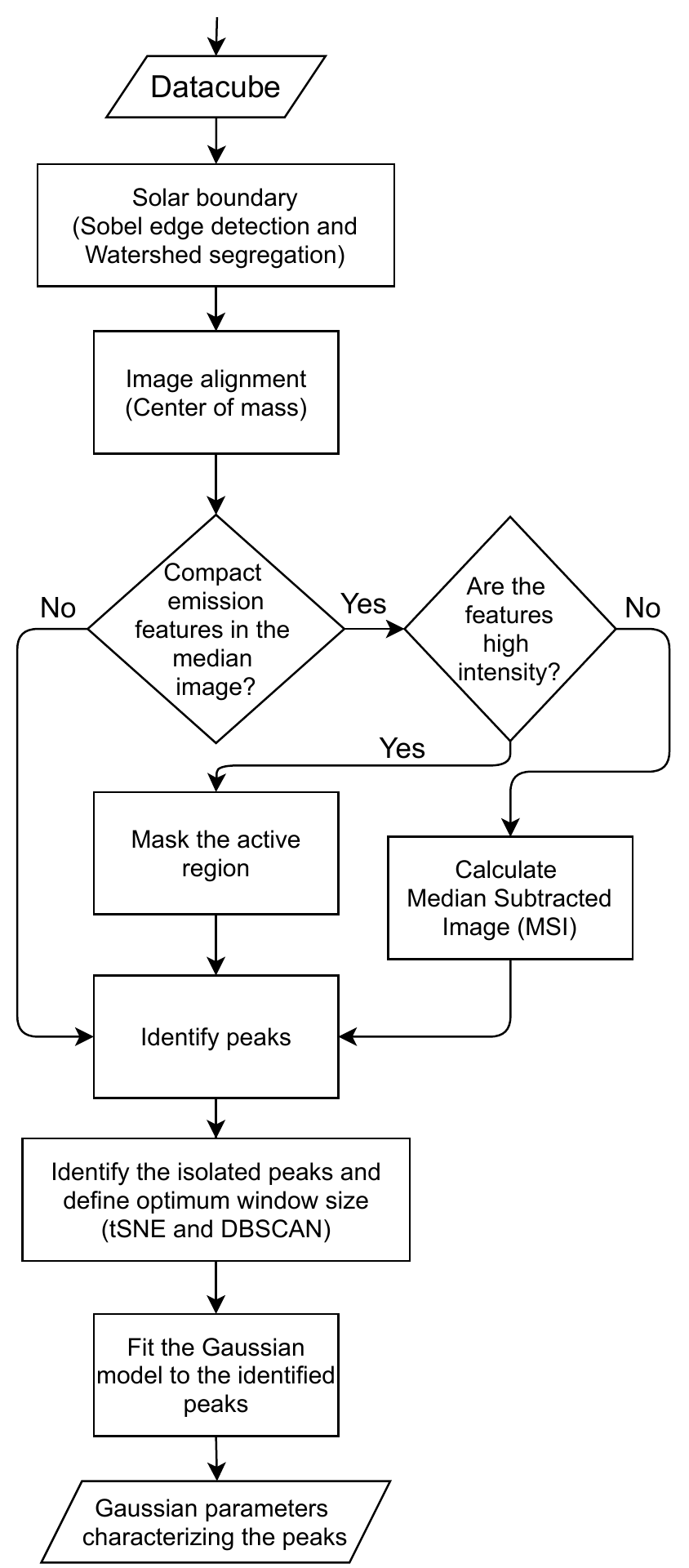}
    \caption{Overall flowchart for identifying and charaterizing the low level peaks seen in the solar radio images.}
    \label{fig:pipeline}
\end{figure}

The input for the pipeline designed for identifying WINQSEs is the 3D image cube $I(x, y, t)$ provided by the radio imaging pipeline as described in the previous section. 
Here $I$ is proportional to the flux density, $x$ and $y$ are the angular coordinates in the images and $t$ runs along the time axis.
At present, we treat the data at different frequencies separately.
The objective of this pipeline is to provide a robust detection and characterization of the low level peaks observed in a given stack of quiet sun radio images. 

This section describes each of the major functional blocks of the pipeline including their necessity, the algorithms of choice and some implementation details. 
Figure \ref{fig:pipeline} gives an overview of the pipeline showing each of the major functional blocks. 
Additional details about the algorithms used are provided in the Appendix. 




\subsection{Determining the Solar Boundary}
\label{subsec:solarboundary}

As evident from Figs.~\ref{fig:solar_image}, the solar radio emission extends well beyond the limb of the Sun.
The emission at metrewave frequencies arises primarily from the million K corona.
As the corona is a diffuse extended region enveloping the Sun, this emission becomes increasingly faint with increasing radial distance and does not have a sharp boundary, in contrast to the optical Sun.
The operational definition of ``boundary of the Sun'', hence, depends on the noise characteristics of the image and inevitably involves some subjective, though well accepted, choices.
In addition, even for the quiet Sun, there are low level but discernible changes in the observed radio emission from one time slice to next.
Therefore it becomes necessary to define the boundary of the Sun independently for each of the images, and this forms the first block of the pipeline.

\begin{figure*}%
    \centering
    {{\includegraphics[width=5.85cm]{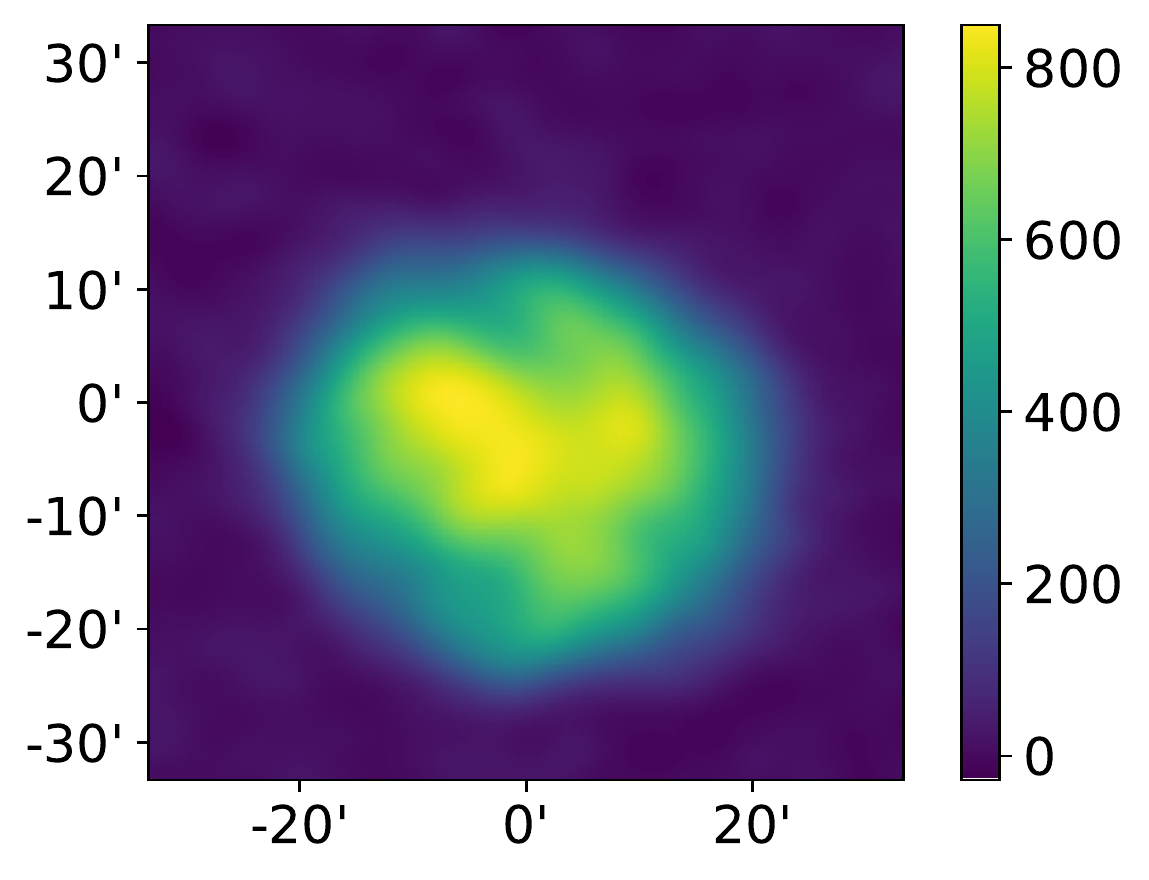} }}%
    {{\includegraphics[width=5.85cm]{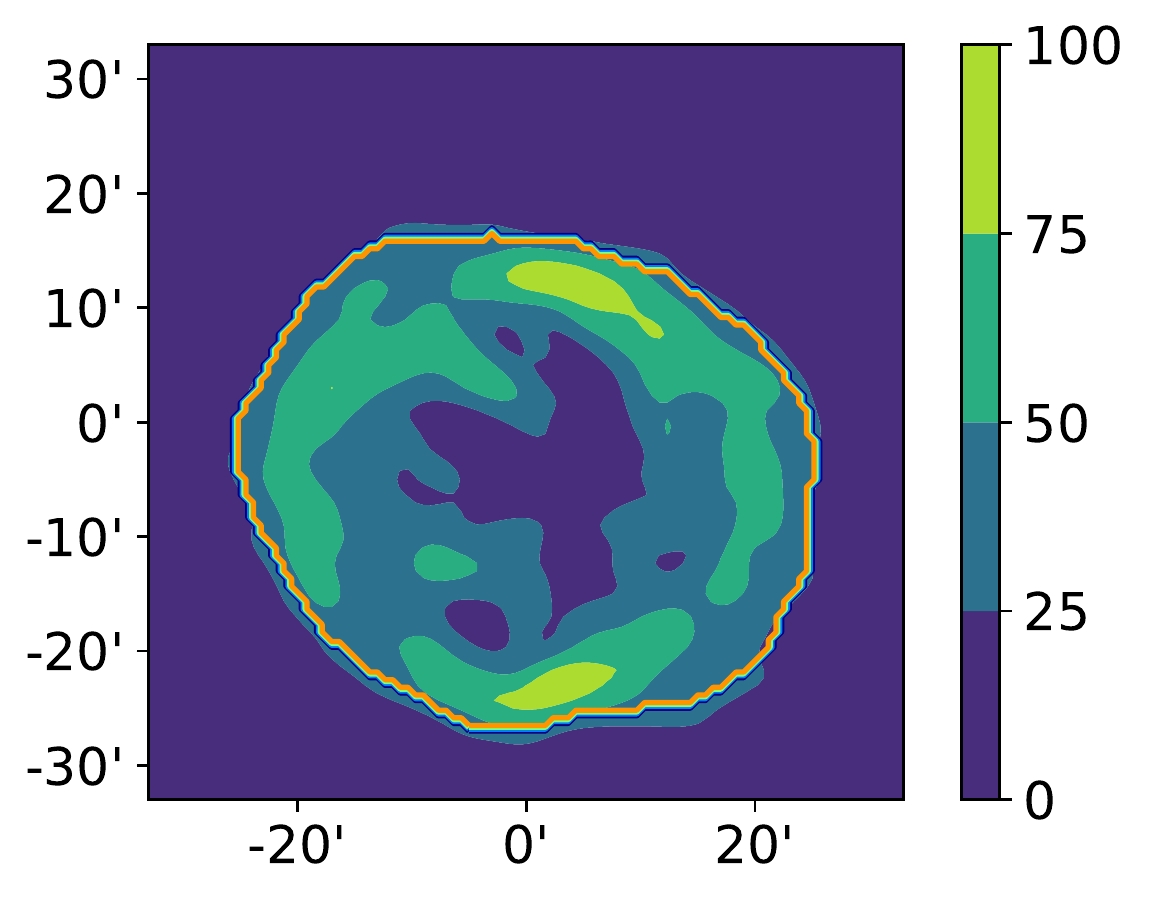} }}%
    {{\includegraphics[width=5.85cm]{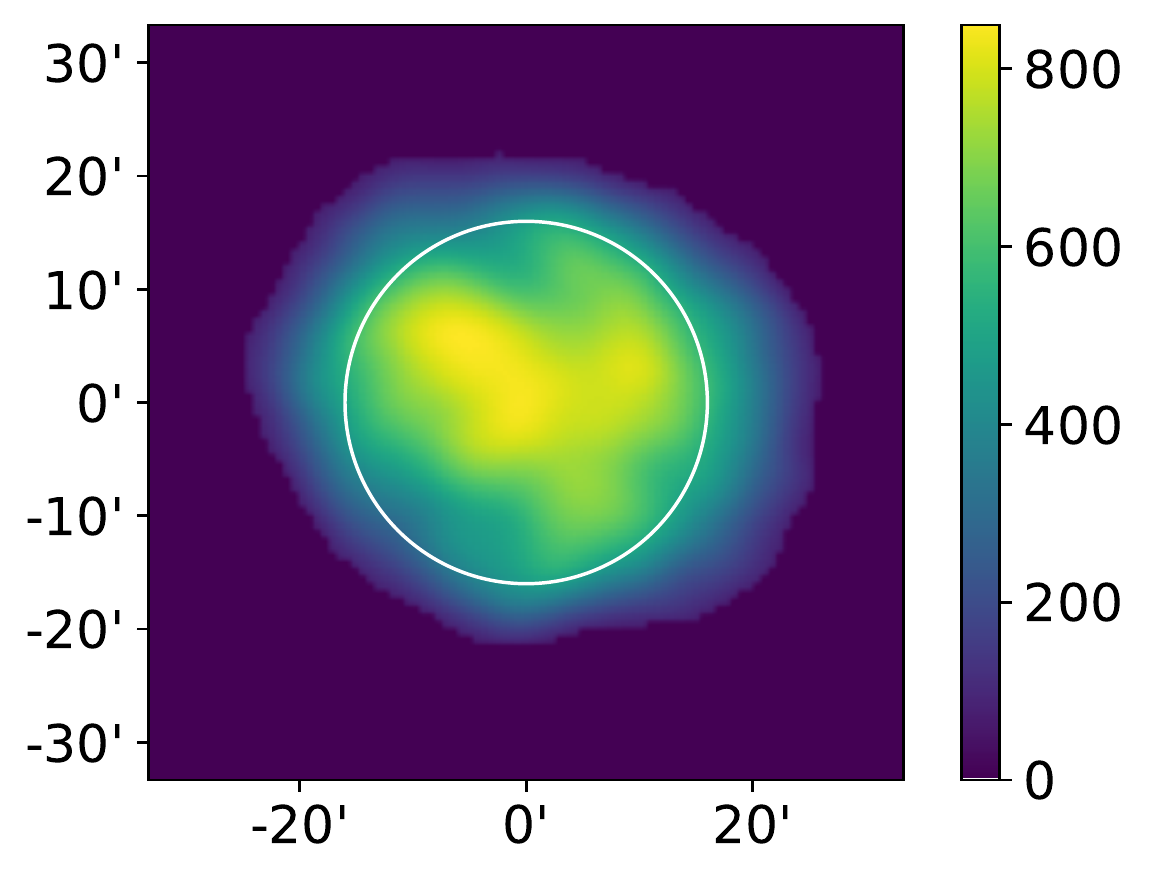} }}%
    \caption{The left panel shows the raw solar image with background noise, also referred to as the elevation map. The middle panel shows the Sobel gradient map for the same image. The boundary of the radio sun (yellow line) is created from this map using Watershed transform to fill the areas of consistently high gradient. The right panel displays the map of the Sun within the boundary defined by Sobel and Watershed methods. This boundary extends well beyond the optical disc of the sun (white circle). This image has also been spatially aligned with other images using the moment of mass method (Sec. \ref{subsec:imageAlignment})}
    \label{fig:solarBoundary}
\end{figure*}

A convenient approach to defining the solar boundary is by applying edge detection techniques to the large intensity gradient observed between the Sun and the region surrounding it.
A region based edge detection algorithm is found to identify the solar boundary of these radio images with sufficient accuracy.
We used the ``elevation map'' (named in analogy with the topographical maps) to define thresholds based on the typical observed intensities well within and well outside the Sun to separate out the pixels that can be unambiguously marked as either the object (the Sun) or the background.
To classify the ambiguous pixels, that could either be the Sun or the background, we employ the following approach.
A Sobel filter is used to define a gradient map of the image.
This filter calculates the gradient in intensity at each pixel of the image in all directions.
An example each of the solar elevation and gradient maps are shown in the left and the middle panels of 
Fig.~\ref{fig:solarBoundary}. 
This clearly shows the high barrier enclosing the solar region in the image.
Next, Watershed object transform, an image  segmentation algorithm \citep{beucher1979watersheds}, is used to fill regions of the elevation map by comparing the gradient at each pixel to that of its surrounding pixels. It starts from the pixels with highest gradient and checks the adjacent pixels to define a path along which the gradient remains high (though not necessarily at maximum). Once it has identified the regions of consistently high gradient, it draws an unbroken line connecting those regions. This process therefore is able to accurately demarcate a continuous border between the sun and the surrounding noise.

This demarcates the radio solar boundary with a high degree of accuracy, as shown in the right panel of Fig.~\ref{fig:solarBoundary}.
This process can also be used to identify boundaries within the sun separating the relatively active, high intensity region from the quiet sun. 
Some more details about the Sobel filter and the Watershed object transform are available in the Appendix.



\subsection{Image Alignment}
\label{subsec:imageAlignment}
Self-calibration algorithms are routinely used in radio interferometric imaging to improve the radio imaging quality from a given dataset \citep[e.g][]{pearson1984, Cornwell1989}. AIRCARS also relies on extensive use of such an algorithm \citep{mondal2019}. A consequence is that the individually self-calibrated images lose positional alignment leading to apparent shifts in the position of the Sun between independently self-calibrated images. The data used here comes from multiple observing scans, each of which were processed independently. In addition, multiple self-calibrations were done as required within individual scans. The objectives of the present study, however, require us to align the solar radio images, and this forms the next block of the pipeline.

\begin{figure}
\centering
    \includegraphics[width=8cm]{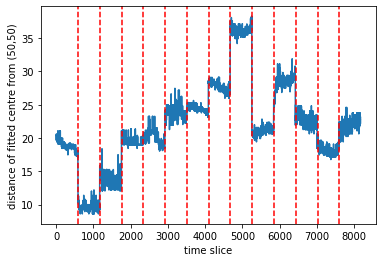}
    \caption{The distance of the centroid of solar emission from a reference pixel for each of the time slices used for 20200620. 
    The red dashed lines represent the boundaries of the adjacent scans. As expected, sudden jumps in the solar centroid location lie at scan boundaries, and the variations inside a scan are smoother and much smaller.
    }
    \label{fig:image_shifts}
\end{figure}

In our exploration, we found that an efficient way to do this, in the present context, is by using the concept of ``centre of mass'' of an image \citep{CHAUDHURI19911}.
The solar boundary determined using the prescription in Sec. \ref{subsec:solarboundary} is used to define a mask with values equal to unity inside the Sun and 0 outside.
The centre of mass, $(h,k)$, of a given image is then defined as:



\begin{equation}
h = \frac{\sum_{i=1}^N{x_i}}{N},\ \ k = \frac{\sum_{i=1}^N{y_i}}{N},
\end{equation}

where $(x_i,y_i)$ are the coordinates of pixels inside the solar mask and $N$ is the total number of pixels inside the solar mask. To align them, the images are shifted to a coordinate frame with $(h,k)$ as the origin, i.e.
\begin{equation}
x' = x-h_r,\ \ y' = y-k_r,
\end{equation}
where $(h_r,k_r)$ are values of $(h,k)$ rounded to the nearest integer and $(x',y')$ are the new pixel coordinates of the aligned images. 
The shifts estimated by this approach are shown in Fig. \ref{fig:image_shifts} in units of pixels.
As expected, the observed shifts change discontinuously at scan boundaries and can be quite significant, when compared to the resolution of the images.



\subsection{Masked or Median Subtracted Images}

\begin{figure}
\centering
    {{\includegraphics[width=4.1cm]{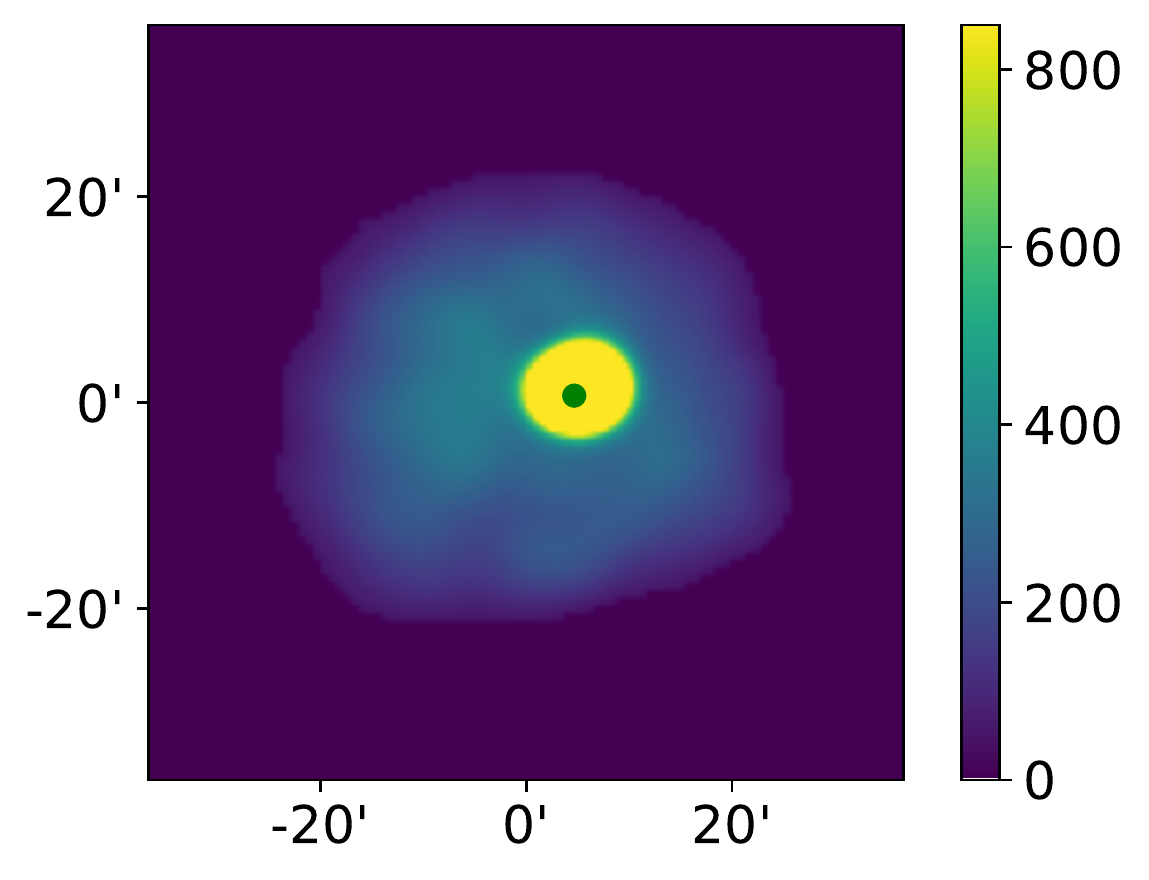} }}
    {{\includegraphics[width=4.1cm]{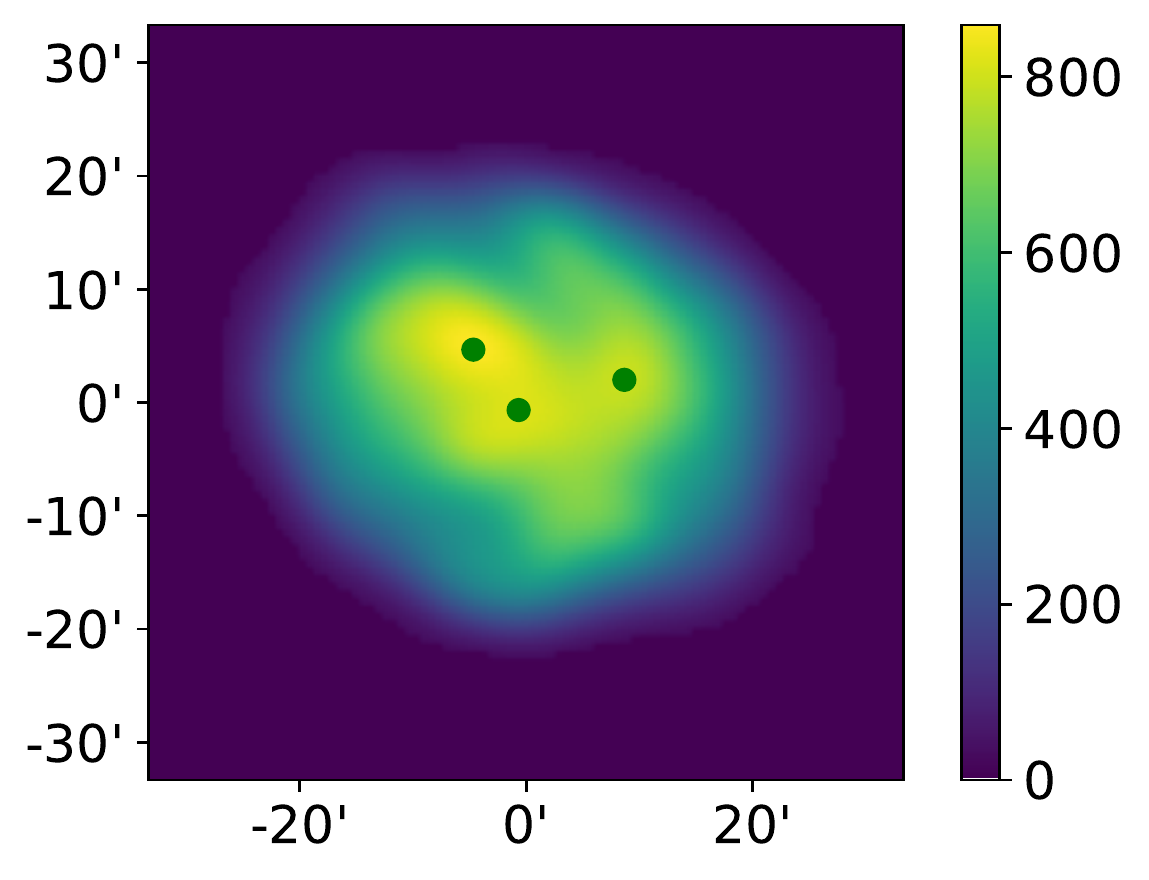} }}%
    \caption{The left panel show the median image for Dataset 20171127, with the active region marked by a green dot. The right panel shows the median image for Dataset 20200620 at 120.52 MHz frequency, with the three persistent peak positions marked with green dots.}
    \label{fig:median}
\end{figure}

The objective of this study is to identify and characterize the morphology of weak short lived sporadic emissions. 
To avoid possible contamination from persistent compact emissions on the Sun, we examined the median images for both these datasets.
The median image, $\tilde{I}(x,y)$, for a given frequency was constructed using the median value of distribution of intensity for every pixel in the time series of aligned solar images. 
$\tilde{I}(x,y)$ for the two datasets are shown in Fig. \ref{fig:median}.
The left panel shows $\tilde{I}(x,y)$ for 20171127 data at 132 MHz and the right panel that for 20200620 data at a representative frequency of 120.52 MHz, the median images for other frequencies are very similar. 
The expectation for 20171127 data, based on M20, was that in the vicinity of the lone active radio emission region (Fig. \ref{fig:median}), $\tilde{I}(x,y)$ will not reflect the quiet Sun emission, but rather the very frequent active emission from this site. 
$\tilde{I}(x,y)$ was indeed in agreement with this expectation. 
The bright compact peak associated with the active emission is much brighter than the quiet Sun regions.
To avoid contamination from this very strong emission, we fit a 2D Gaussian to the active emission and mask a region corresponding to 2.5$\sigma$ centered on the peak of the best fit Gaussian.

\begin{figure*}
    \centering
    {{\includegraphics[width=5.85cm]{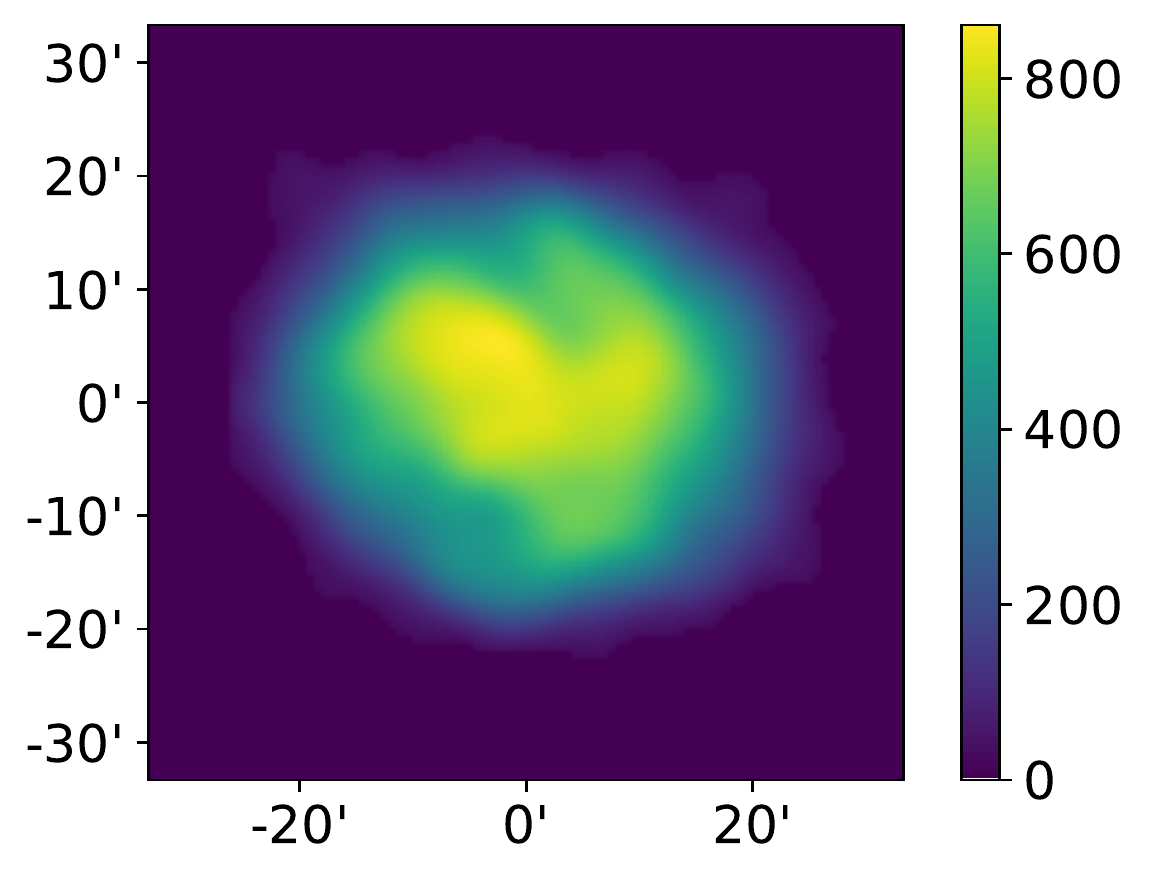} }}
    {{\includegraphics[width=5.85cm]{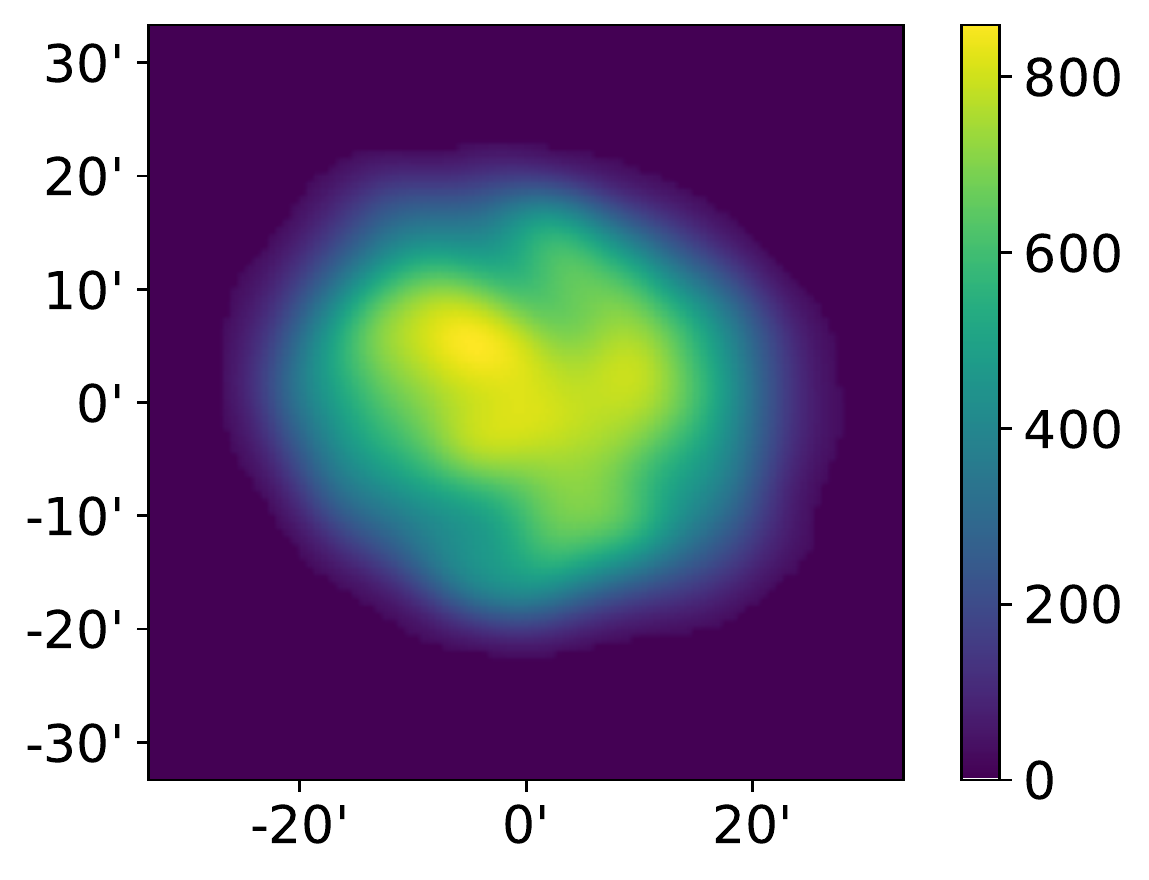} }}
    {{\includegraphics[width=5.85cm]{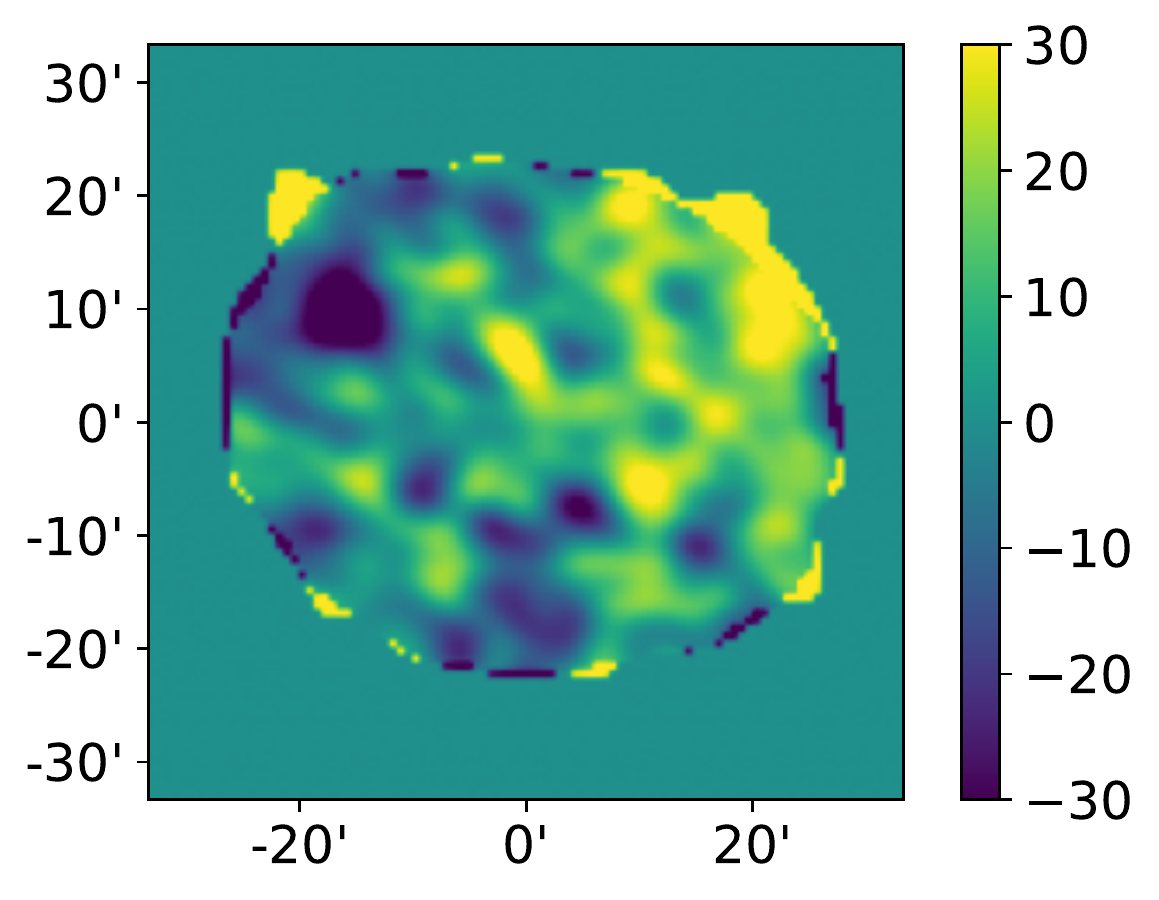} }}
    \caption{The left panel shows the solar image at a particular time  for the  Dataset 20200620 at 120.52 MHz, the middle panel shows the median image over time for the entire dataset, while the right panel shows the Median Subtracted Image (MSI) for the same time slice as shown in the left panel. 
    }
    \label{fig:MedianMap}
\end{figure*}

For the 20200620 dataset, $\tilde{I}(x,y)$ was expected to represent spatially smooth quiet Sun emission. 
The weak intermittent features of interest here were expected to be infrequent enough to not influence the median image.
While the bulk of the image is spatially smooth, it does show three weak compact emission peaks. 
For this dataset, to avoid contamination from these weak compact emissions, we subtract $\tilde{I}(x,y)$ from the images for individual time slices, $I(x,y,t)$, and work with these median subtracted images, $MSI_t$, in the downstream sections of the pipeline. As an illustration, in figure~\ref{fig:MedianMap}, the left panel shows the solar image at a particular instant for the frequency 120.52 MHz, the middle panel shows the median image over the entire time, and the right panel shows the median subtracted image corresponding to the left panel, which is used in further analysis.

{\color{blue} 
}


\subsection{Identifying Feature Peaks}
\label{subsec:peakIdentification}
\begin{figure}
    \centering
    \includegraphics[width=9cm]{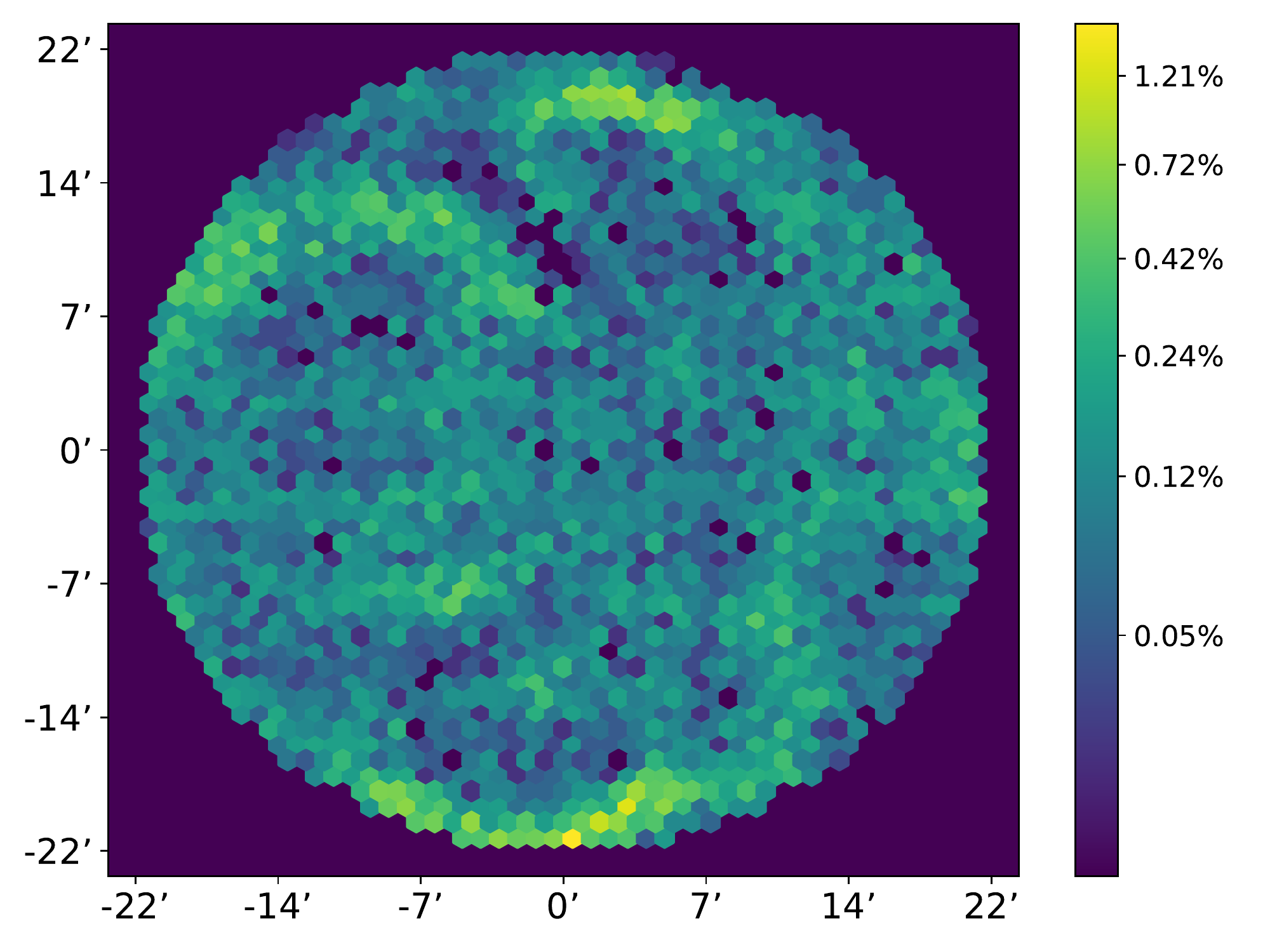}
    \caption{Distribution of the WINQSEs detected in a region of diameter 40' centered on the sun 
    for Dataset 20200620 at 120.52 MHz.
    }
    \label{fig:incidence}
\end{figure}

A prerequisite for characterising the morphology of these weak emissions was identifying the location and position of the peak features in the image.
A threshold of $5 \sigma$ was used to select features with sufficient signal-to-noise, where $\sigma$ is the RMS noise in the map away from the Sun.
A pixel, $(x_p,y_p)$, was considered to be a reliably detected feature if the intensity of all other pixels in a $N \times N$ kernel centered on $(x_p,y_p)$ were lower than $S(x_p,y_p)$, the intensity at $(x_p,y_p)$, where $N$ is the size of the kernel in pixel units. 
The size of the kernel for feature detection needs to be large enough to reject the small scale noise fluctuations, and compact enough to not be affected by other nearby peaks. 
On trying different kernel sizes, we found that the $5\times5$ kernel works best in identifying true features as opposed to noise in both the datasets. 
Since the size of the PSF were 5.6 and 7 pixels for the first and second datasets respectively, a kernel slightly smaller than these values was well suited for accurately detecting the features.
The distribution of features in the plane of the sky for dataset 20200620 for 120.52 MHz is shown in Fig.~\ref{fig:incidence} over a region 40' in diameter and centered on the Sun in units of percentage of time slices in which these emissions are detected.
Those for other frequencies of this dataset are very similar.

\subsection{Isolated Features and Optimum Fitting Window}
\label{subsec:optWindowSize}


\begin{figure*}
\centering
    \includegraphics[width=14.cm]{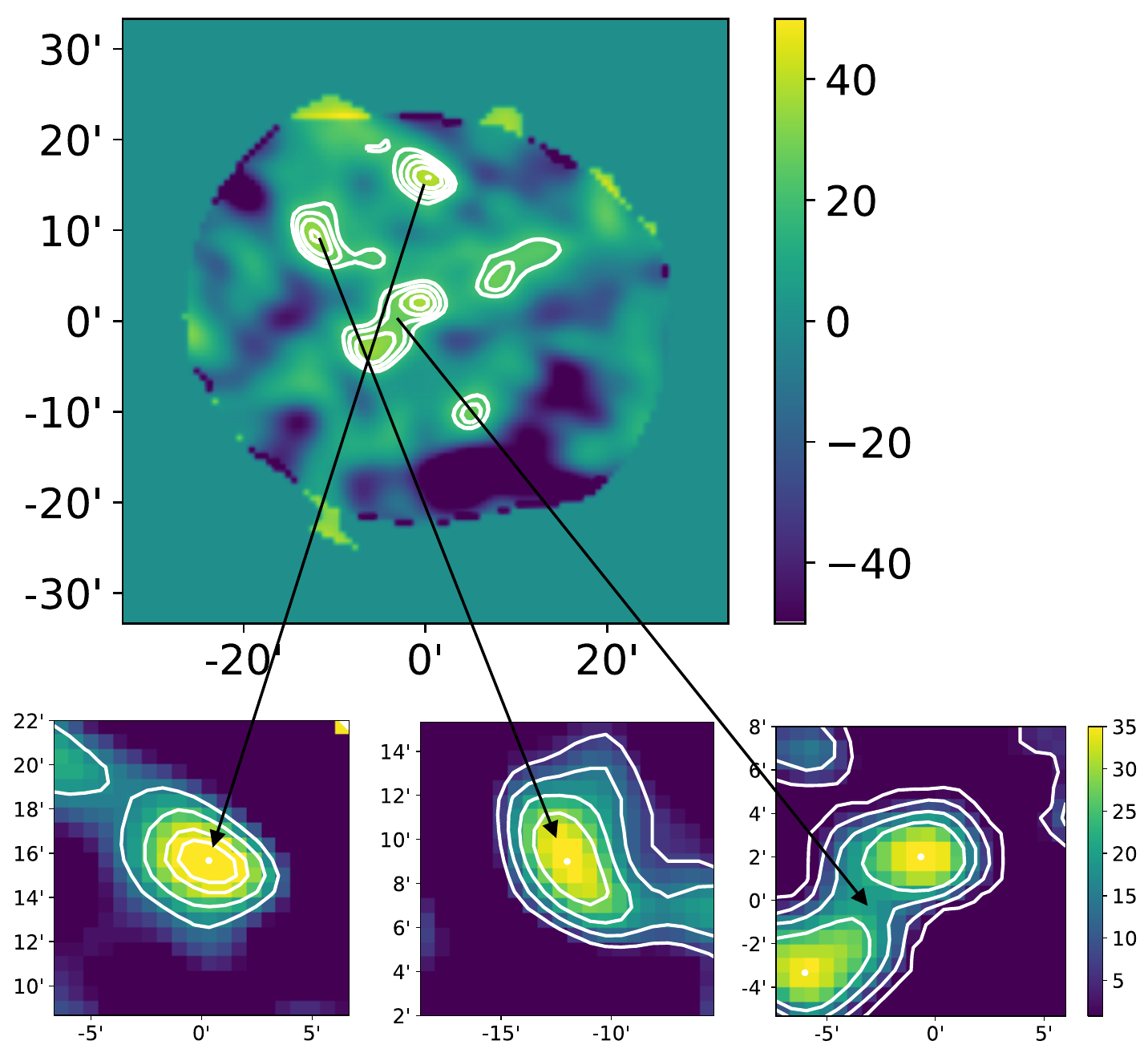}
    \caption{Top panel shows the contours around the features detected at a particular time for the median subtracted 120.52 MHz quiet sun dataset. The bottom panel shows four of these features in detail. We can see that the first feature (left panel) is quite isolated, the second feature (middle panel) is semi-isolated, with some distortion in contours due to the proximity of nearby features, and the third and fourth features (right panel), are so close together that the contours around them are joined.}
    \label{fig:contours_msi}
\end{figure*}

Next we examine the intensity variations observed around the features detected to understand their nature. 
We find that the features can be broadly classified into three types. Some features are isolated, with no nearby feature strong enough to impact their intensity distribution. 
Some features are semi-isolated, where the intensity distribution around a given feature is somewhat distorted due to the contribution from other nearby features. 
The third kind are the clustered features, where two or more features are close enough that it is not straightforward to decompose these into individual uni-modal compact emission features.
Examples of these different types are shown in the Fig.~\ref{fig:contours_msi}. 

To investigate the morphology of these features, it is most convenient to work with isolated and semi-isolated features, since these could be described well by a 2D Gaussian model.
The numbers of such features run into tens of thousands for each spectral slices \citep{mondal2020}.
This led to the requirement building an algorithm for automated identification of isolated and semi-isolated features, suitable for characterization using a Gaussian model.
As the quality and robustness of the fit is closely tied to the choice of the window over which the fit is performed, we also needed to devise an algorithm to define the optimal window for this fit.
The choice of an optimal window has to balance opposing considerations. 
The larger the fitting window, the more data is available to constrain the fit, but the possibility of contamination from neighboring features affecting the quality on the fit is also higher. On the other hand, if the window is too small, one does not have enough data to constrain the fit well. 
In our experimentation with the different choices for fitting windows for features with significant ellipticity, use of rectangular windows roughly matched to the feature size and oriented similar to the feature produced the best results.

To identify the different types of features and to find an optimal window for the fit, we first characterized each feature using the detailed characteristics observed in their intensity distribution.
These included the intensity of each feature, the size and orientation of the largest closed contour around it, the drop in intensity from the peak of the feature to the farthest and nearest ends of the largest closed contour around it, and the distance to and the intensity of the nearest neighbouring feature. 
It is reasonable to expect these characteristics to carry the information about isolated or clustered nature of the features.
However, the relative importance of these different attributes in contributing to identifying isolated or semi-isolated features are not known a priori and are not easy to establish.
We found, for instance, that it was not possible to identify the isolated or clustered features reliably based only on their distance from their neighbours. 
Two features could be quite far from each other, but if one was much brighter, it could still distort the intensity distribution in the neighborhood of the other. Similarly, two comparatively nearby features, if compact enough, could still be effectively isolated from each other. 
Therefore we needed to use a judicious combination of all these characteristics to differentiate between features of these types.
Large numbers of prior examples of different types of features, which could potentially be used to pose the task at hand as a classification problem, were not available to us.
Hence, we posed this as an unsupervised clustering problem with the objective of classifying the features. Our rational behind the choice of algorithm for this clustering is further explained in section~\ref{sec:clust}.

To simplify matters, we condensed the information from all the features into two dimensions using t-distributed Stochastic Neighbor Embedding \citep[t-SNE,][]{Hinton_Roweis_2003, vanDerMaaten2008}. 
This algorithm embeds each multi-attribute feature in a two-dimensional phase space such that there is a high probability of similar features mapping to nearby points in this plane while dissimilar ones are separated by larger distances.
This considerably reduced the complexity for the next step which used an unsupervised clustering algorithm.

\begin{figure}
\centering
    \includegraphics[width=8.5cm]{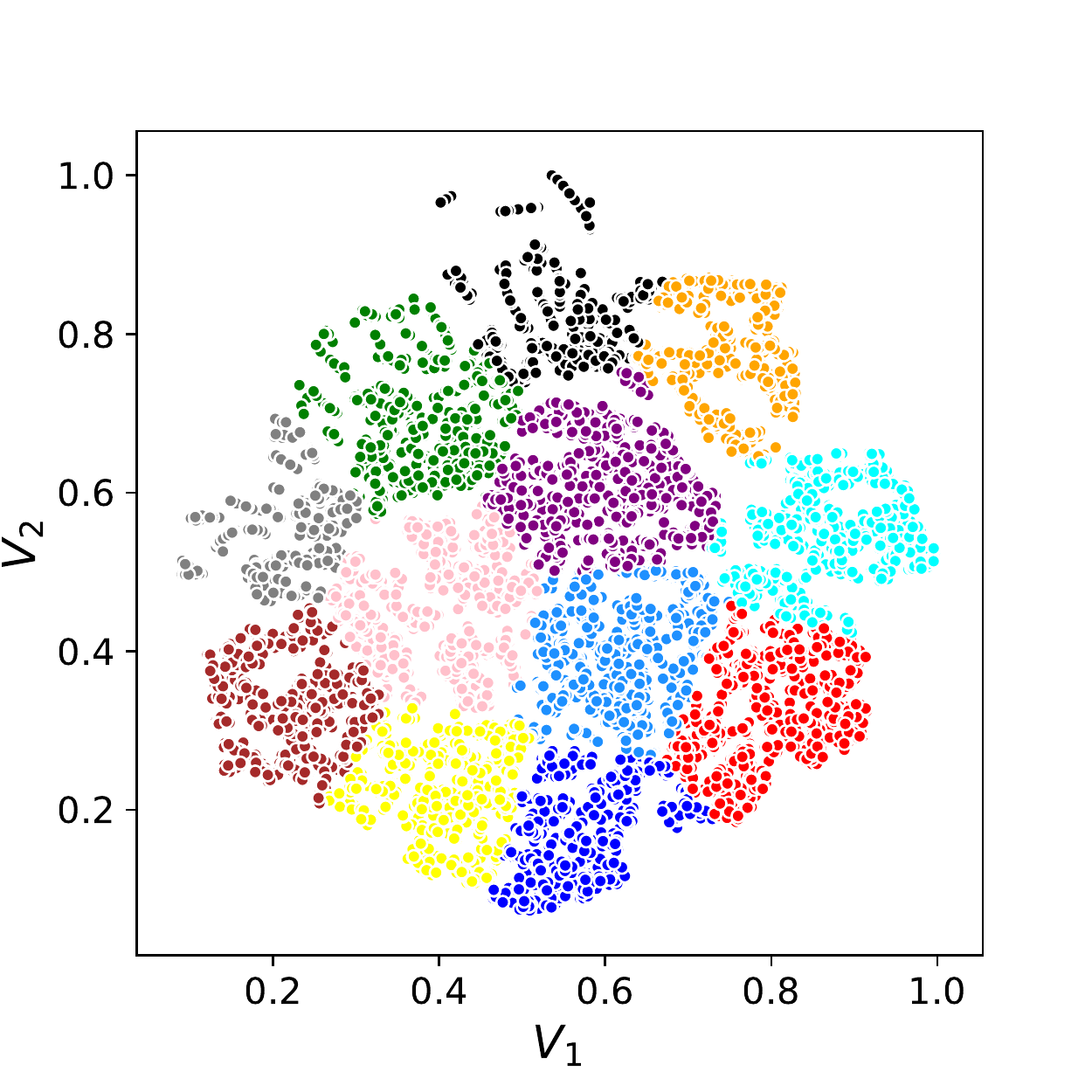}
    \caption{Unsupervised clustering of the identified features using t-SNE+DBSCAN, for the 120.52 MHz quiet sun dataset, represented in the t-SNE variables $V_1, V_2$. Each colour represents a group of features which have been found to have similar properties by the algorithm. We chose three of these groups, Group 1 (blue), Group 2 (orange), Group 3 (gray), for further study.
    }
    \label{fig:groups}
\end{figure}

\begin{figure*}%
    \centering
    {{\includegraphics[width=5.3cm]{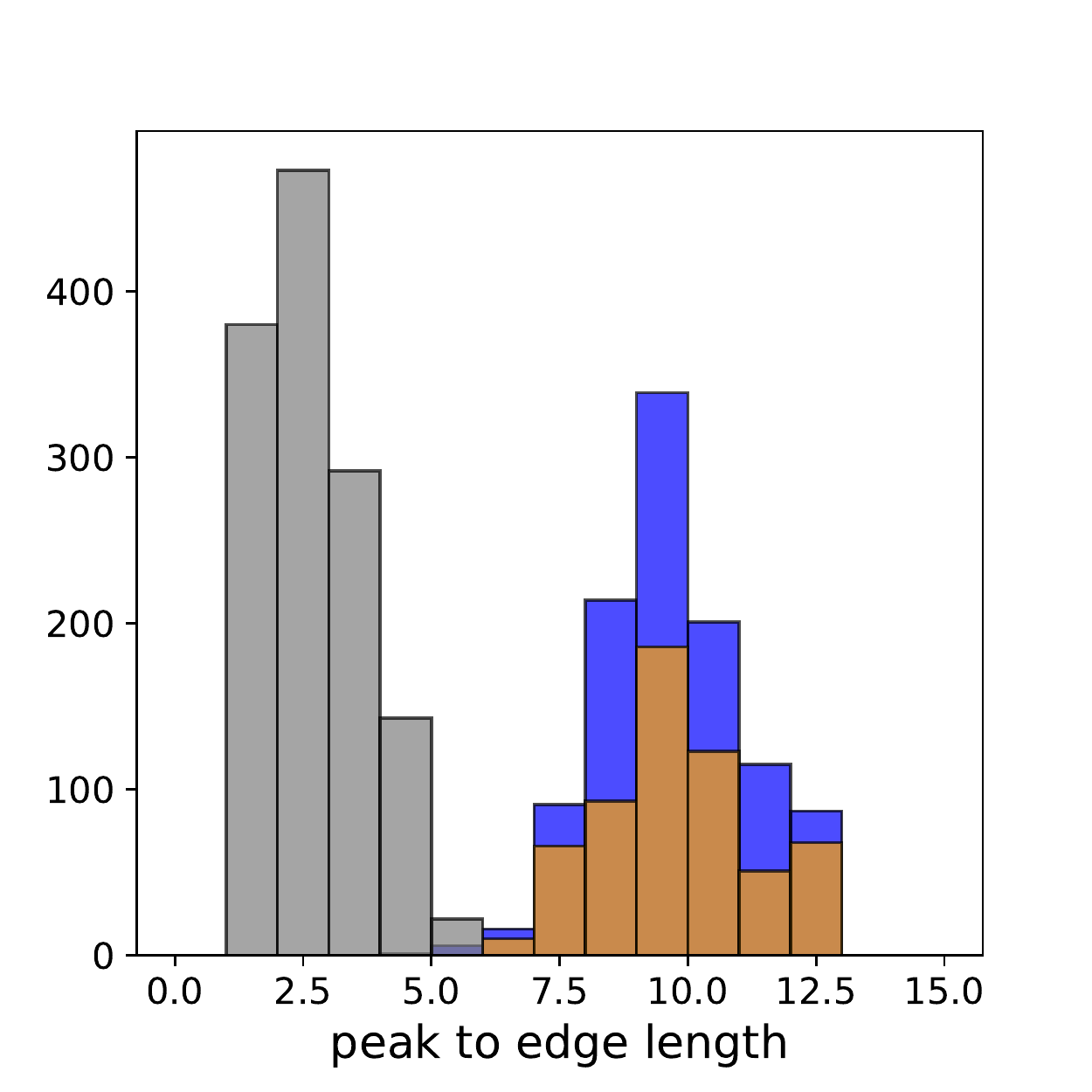} }}%
    {{\includegraphics[width=5.3cm]{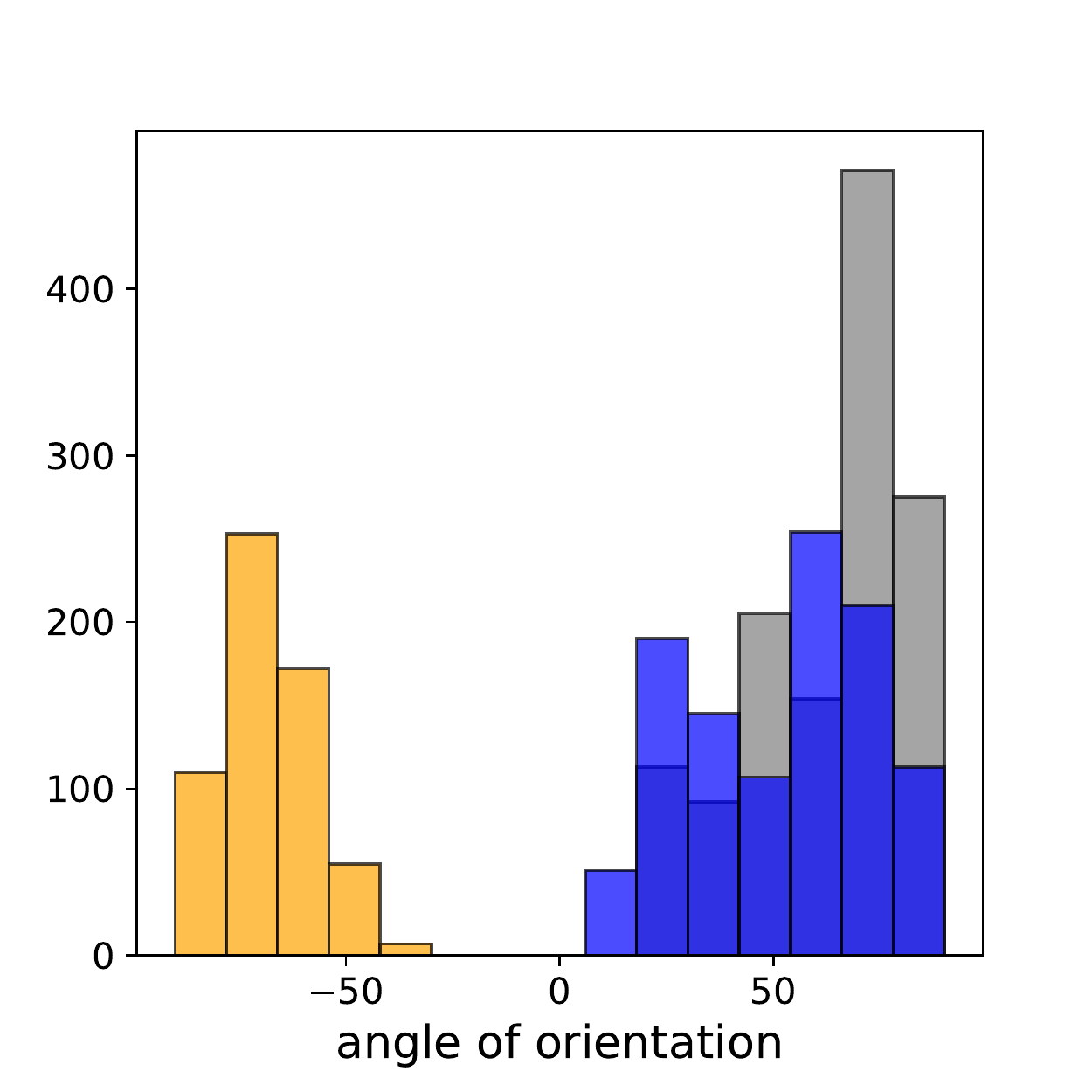} }}%
    {{\includegraphics[width=5.3cm]{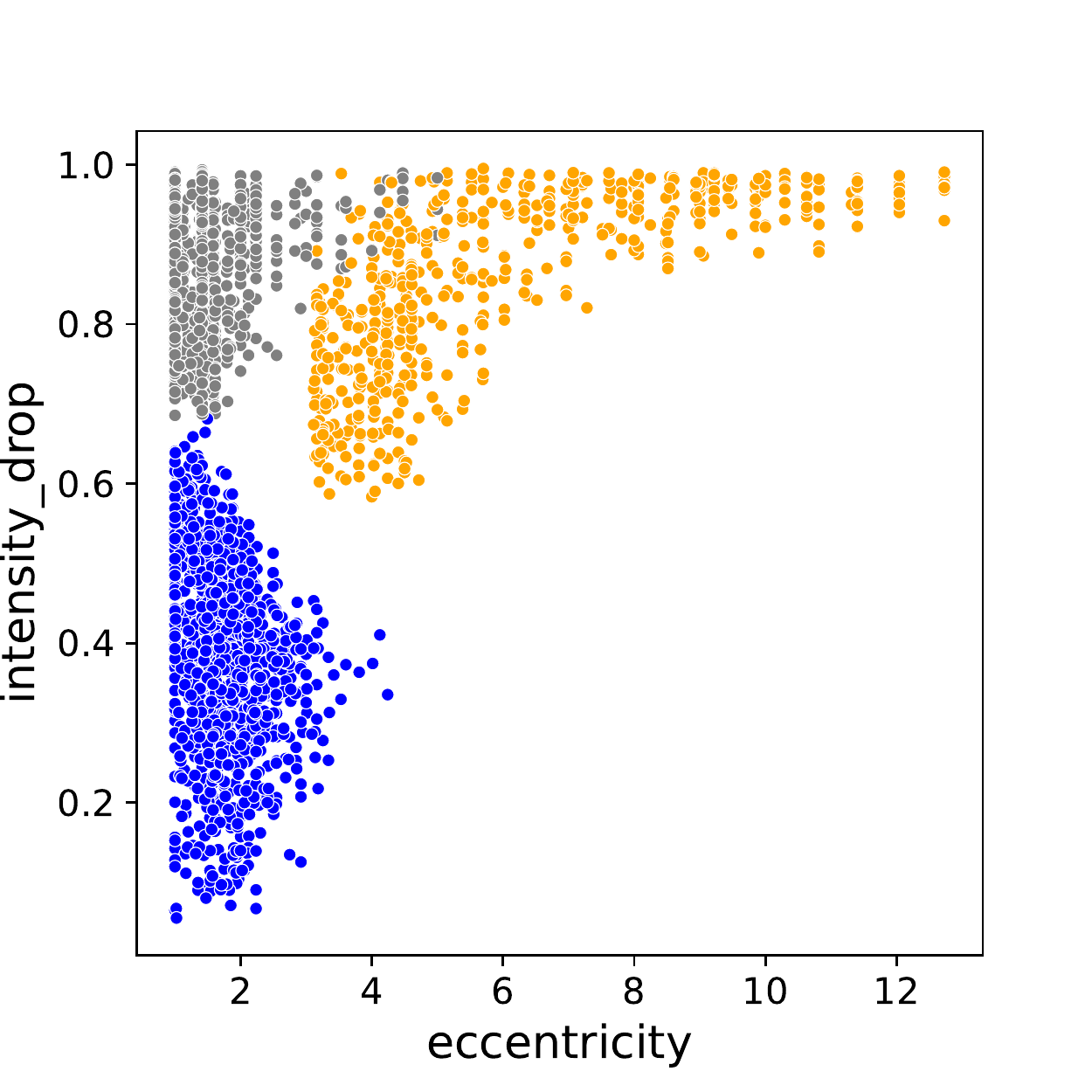} }}%
    \caption{Distribution of various attributes for three chosen groups from Fig.~\ref{fig:groups}. The left panel shows the histograms for the maximum distance between the feature peak in a given cluster to the largest closed contour around it. The middle panel shows the histograms for the angle of orientation of the largest closed contour around the feature. The right panel shows the distributions in the eccentricity of the largest closed contours against the drop in intensity from the feature to the farthest edge of the contour around it.}
    \label{fig:groups_params}
\end{figure*}

Density-Based Spatial Clustering of Applications with Noise \citep[DBSCAN,][]{ester1996density} is utilized next to group similar features together, and to identify outlier features. 
This algorithm does not require the number of groups to be specified at the outset, it rather calculates the optimum number of groups based on the distance between the different points in the t-SNE space.
We found that between nine and thirteen groups were forming for the different datasets. 
An example of the groups thus formed is shown in Fig.~\ref{fig:groups}. 
This particular dataset was separated out into twelve 
different groups, represented by different colours. 
To illustrate the differences between the characteristics of these groups we examine three of them in greater detail -- 
Group 1 (blue), Group 2 (orange), and Group 3 (gray). 
These three groups were chosen as they were well separated from each other in the t-SNE space. 
Some details of the characteristics of the features of each groups are shown in Fig.~\ref{fig:groups_params}. 
One of the characteristics examined was the maximum distance from the feature peak to the edge of the largest closed contour encircling it.
This is indicative of how compact the feature is. 
Groups 1 and 2 were both found to contain comparatively larger features, while the Group 3 features were found to be compact. 
The angle of orientation of the largest closed contour around the features was also examined. 
Groups 2 and 3 were found to occupy essentially the same range of positive inclinations from $\sim$5$^{\circ}$ to 90$^{\circ}$, while Group 1 was found to occupy the negative inclination range from $\sim$-80$^{\circ}$ to -40$^{\circ}$.
The eccentricities of the features 
and the drop in intensity from the feature peaks to the edge of the contour were also examined.
Group 1 was found to have lower eccentricities, and larger drops in the intensity; Group 2, higher eccentricities and a comparatively lower drop in intensity; while Group 3 had low eccentricities and lower fall in intensity.
From these characteristics, it is evident that Group 3 contained the most compact features, but these features also showed low variation in intensity. Group 1 contained larger features, with low eccentricity, and higher fall in intensity from feature to edge of contour. Group 2 contained larger, highly eccentric features, with low variation of intensity. It was therefore expected that Group 2 would be most likely to contain the clustered features, hence the larger sizes and eccentricities of contours with low intensity variation. Groups 1 and 3 most likely contained the isolated or semi-isolated features. 
An examination of some individual features each of these groups supported this hypothesis. 
Similar investigations for other groups for other frequencies of this Dataset yielded similar results. 

The t-SNE+DBSCAN algorithm appeared to separate out the features using a combination of all the attributes available, rather than by individual attributes, and in doing so, was found to accurately separate the isolated, semi-isolated and clustered features.
The separation into groups had the added advantage that we could now define { well-informed} window sizes for each group of features. We used the median size and orientation of the largest contour around the feature for each group to define the optimal window over which the fitting was done. 
The use of median sizes is reasonable because, within a group, the distribution for the attributes such as the angle of orientation etc. are quite compact as compared to the entire range of possible values. 
We have also verified that defining a median window for each group of features (each containing around a thousand features) gives results comparable to fitting each feature with a window specific to its individual characteristics.
Some more details about the t-SNE and DBSCAN are provided in the Appendix.

\subsection{Fitting a Gaussian model to the peaks}
\label{subsec:fitGaussian}

Having identified the features of interest (isolated and semi-isolated features) and estimated the windows over which to fit the 2D generalized Gaussian models for them, the next step is to determine the best fit model parameters.
The 2D Gaussian, { $G$}, is given by
\begin{eqnarray}\label{eqn:gaussianModel}
\phi_x &=& \frac{(x - x_0)\cos \theta - (y-y_0)\sin \theta}{\sigma_x} \\
\phi_y &=& \frac{(y - y_0)\cos \theta + (x-x_0)\sin \theta}{\sigma_y} \\
G &=& O + A \frac{2}{\pi^2 \sigma_x \sigma_y} e^{-\frac{1}{2}[\phi_x^2 + \phi_y^2]} \,\,,
\end{eqnarray}
where,
\begin{conditions}
O & offset\\
A & the Gaussian amplitude\\
x_0, y_0 & position of the Gaussian feature\\
\sigma_x, \sigma_y & half widths of the major and the minor axes\\
\theta & angle of orientation w.r.t the x axis.
\end{conditions}

\begin{figure*}
    \centering
    \includegraphics[scale=0.5]{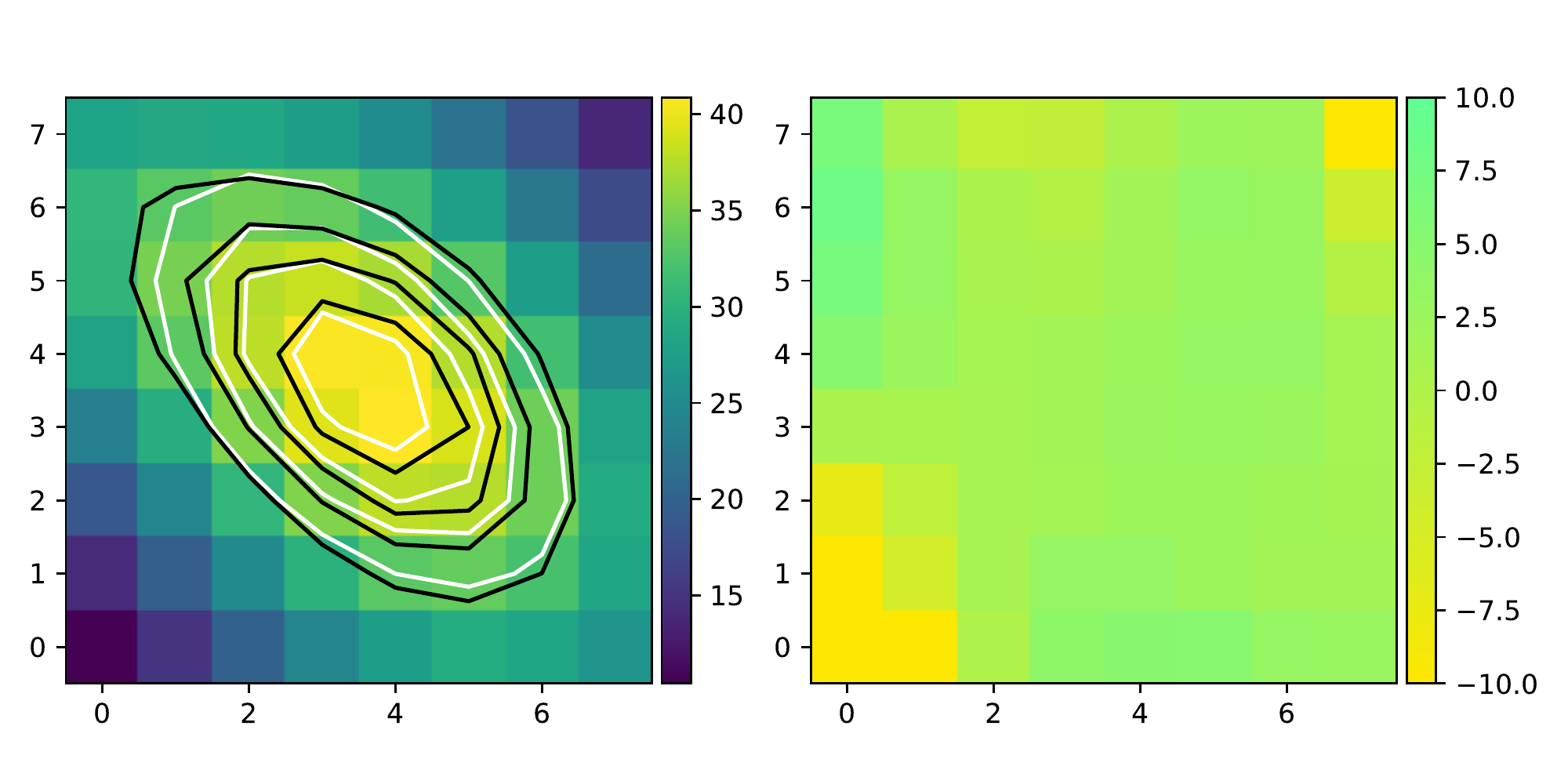}
    \includegraphics[scale=0.4]{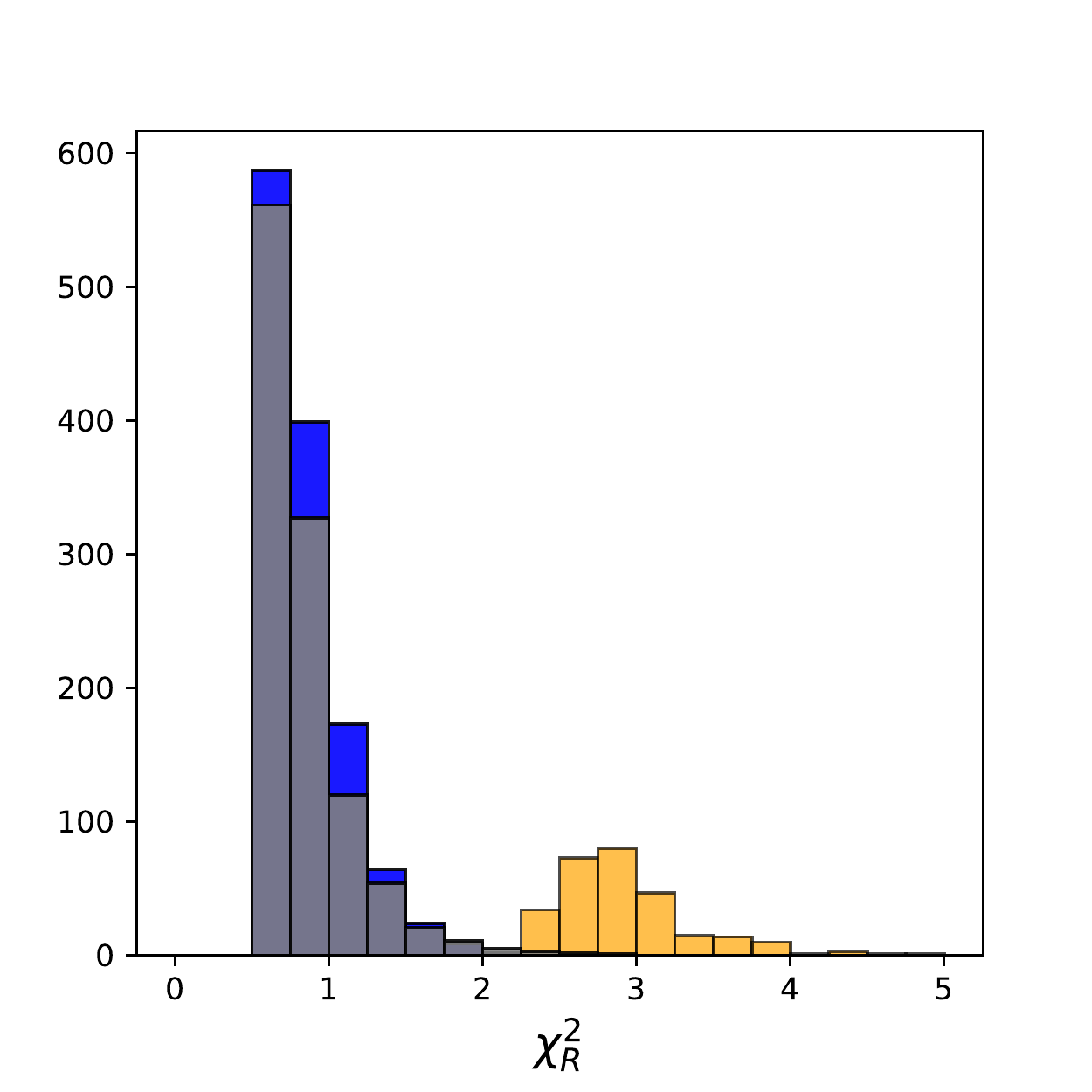}
    \caption{The figure shows a zoomed in view of an example WINQSE to demonstrate the quality of Gaussian fit. 
    The left panel shows the contours of the actual image in black and those corresponding to the best fit Gaussian model white. The middle panel shows the residual of the Gaussian fit. 
    The contours mark 78.3\%, 85.6\%, 90.5\%, 95.4\% of the peak.
    The residuals lie in the range $\pm$5\% in the regions where the Gaussian has a significant amplitude and increase to $\pm$10\% towards the edges of the image. Right panel: Distribution of the reduced $\chi^2$ for peaks in the three chosen groups from Fig.~\ref{fig:groups}.
    }
    \label{fig:gaussianFits}
\end{figure*}

Figure \ref{fig:gaussianFits} shows an example Gaussian fit to an isolated feature in the left panel and the corresponding residual in the right panel.
After the fitting, we further filtered out features whose $\chi^2$ values and errors on the individual parameters were greater than $Q3 + 1.5\ IQR$, where $Q3$ is the 3rd quartile and $IQR$ the inter quartile range for the $\chi^2$ and error distributions. {$\chi^2$ is defined as

\begin{equation}
    \chi^2=\Sigma_{i=0}^{i=N} \left (\frac{O_i-M_i}{E_i}\right)^2
\end{equation}
Here $(O_i,E_i,M_i)$ are the observed value, its associated error and the corresponding value predicted from the obtained Gaussian fits. }
This {filtering scheme} ensured that fits with high errors on the parameters, or poor overall $\chi^2$, are not considered for further analysis. 
In the right panel of Fig.~\ref{fig:gaussianFits}, we show the distribution of the reduced $\chi^2$ values for the three groups which had been considered in Sec. \ref{subsec:optWindowSize}.
{Reduced $\chi^2$ is defined as the ratio of $\chi^2$ to the number of degrees of freedom, where the number of degrees of freedom is defined as $D-N$, with D and N being the total number of data points and free parameters respectively.}
Group 1 was found to have $1303$ features, of which $1258$ (96.5\%) were fit well, Group 2 had $1507$ features, of which only $281$ (18.6\%) were properly fit, while Group 3 had $1156$ features of which $1105$ (95.6\%) were good fits. 
For Group 2, even the few features that were fit tended to have higher $\chi^2$. This demonstrated that typically the clustered features had higher residuals and most isolated and semi-isolated features were fit well by a 2D Gaussian, indicating that our clustering algorithm was successful in grouping the features and finding the optimal fitting windows for them. Thus our approach of arranging the features into groups by their characteristics and fitting them optimally in those groups is able to automate the otherwise laborious process of separating out the isolated features and individually defining optimal fitting windows in order to fit them with Gaussians.

\section{Results}
\label{sec:results}
This section presents the results from Gaussian fitting procedure just described.

\subsection{Spatial distribution of WINQSEs}
Figure~\ref{fig:incidence} shows the fractional occupancy of peaks of emissions at different locations on the Sun. The fractional occupancy is defined as the fraction of observation time for which WINQSEs were detected at any location on the Sun. 
The radio Sun is larger in size than the 40' diameter region shown in Fig. \ref{fig:incidence}.
However close to its edge, there is an increased prevalence of spurious features in the median subtracted images used here.
To avoid contamination from such features, we restrict the region examined to a diameter of 40', well inside the edge of the median image.
It is evident that these features are ubiquitous and their distribution varies smoothly with position. 

\subsection{Distribution of intensities of WINQSEs}

\begin{figure*}
    \centering
    \includegraphics[width=18.5cm]{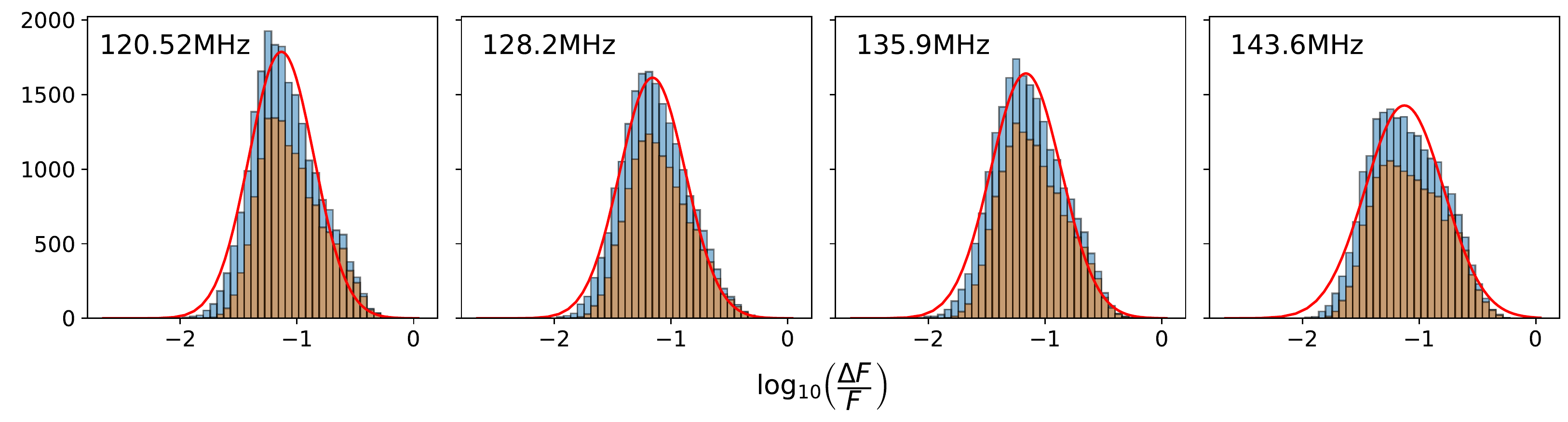}
    \caption{Distribution of $\Delta F/F$ for WINQSEs detected within the solar boundary for Dataset 20200620 at 120.52 MHz, 128.2 MHz, 135.9 MHz and 143.6MHz. 
}
    \label{fig:deltaFbyF}
\end{figure*}

The identification of the peaks of these features, the spatial distribution of which is shown in Fig. \ref{fig:incidence}, also enables us to examine the distribution of their strengths.
In analogy with M20, we define a quantity $\Delta F(x,y)$ as: 
\begin{equation}
    \Delta F(x,y)_t = F(x,y)_t - \tilde{F}(x,y),
\end{equation}
where $\tilde{F}(x,y)$ is the median image for a given frequency and $F(x,y)_t$, the image at time $t$ for the same frequency.
Histograms of $\Delta F(x,y)/\tilde{F}(x,y)$ for all time slices for a given frequency are shown in Fig. \ref{fig:deltaFbyF} for each of the four frequencies analyzed here.
The histograms in blue show the distribution for all reliably detected peaks. 
Those in orange correspond to the subset deemed to be well fit by a Gaussian model and are discussed in Sec. \ref{sec:fits}. 
The red line shows the lognormal fit to the blue histogram. It is evident from this figure that the lognormal distribution is a good representation of the observed distribution. 

\subsection{Gaussian fitting results}
\label{sec:fits}

\begin{figure}
    \centering
    \includegraphics[width=9cm]{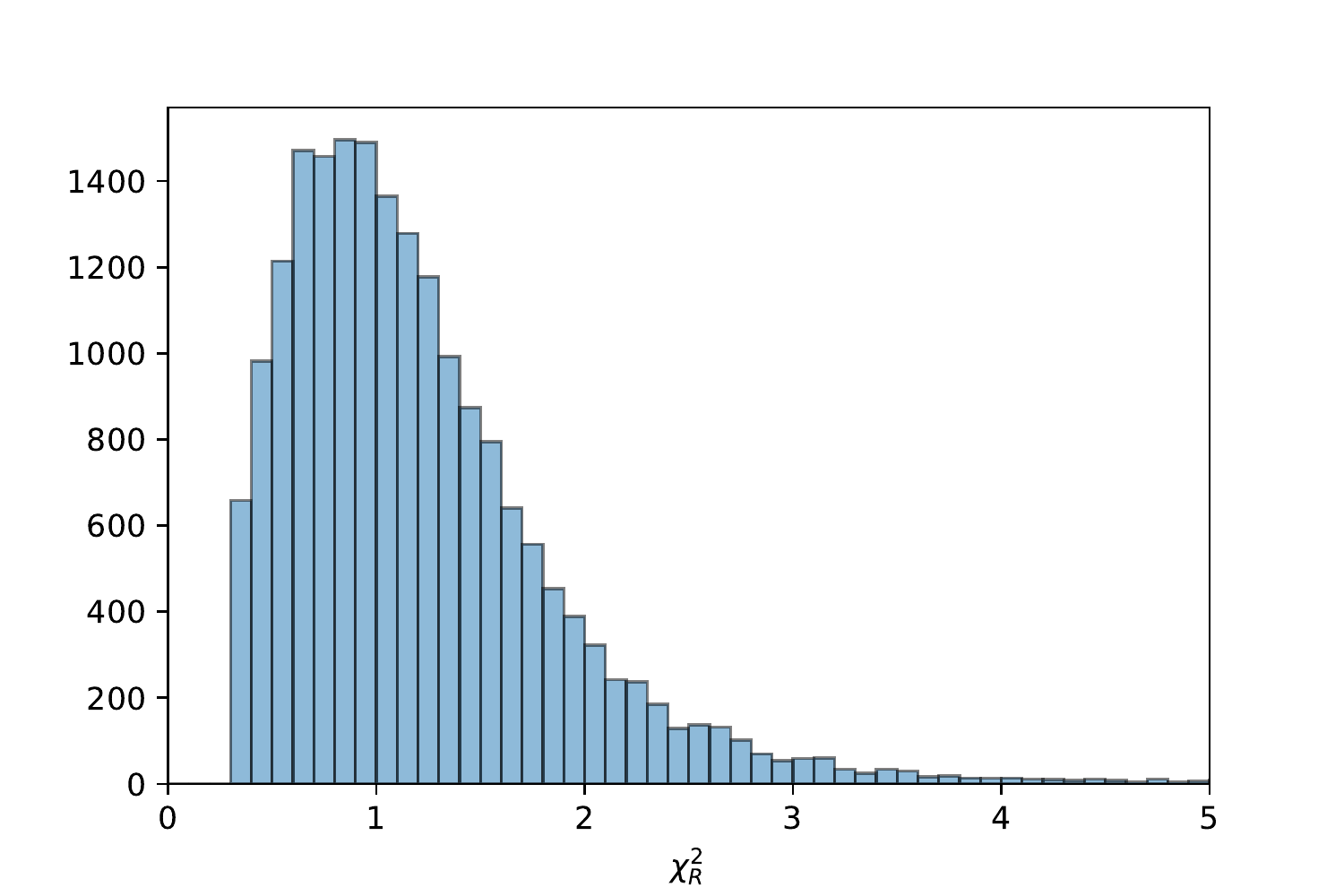}
    \caption{Distribution of the reduced $\chi^2$ for WINQSEs detected within the solar boundary for Dataset 20200620 at 120.52 MHz.}
    \label{fig:chi2}
\end{figure}

The results from the Dataset 20200620 at 120.52 MHz are described in detail. Unless otherwise mentioned, the results at other frequencies are very similar.
Figure~\ref{fig:chi2} shows the reduced $\chi^2$ distribution of the fitted features detected in a region of 40' diameter centered on the sun{, after they have been filtered as described in section~\ref{subsec:fitGaussian}}.
It is clear that the vast majority of the features have a reduced $\chi^2$ close to 1, implying that the chosen Gaussian model represents a good fit to the data.
Henceforth, the features which are fit well by 2D Gaussians are referred to as detected WINQSEs. 

\begin{figure*}
    \centering
    {{\includegraphics[width=14.5cm]{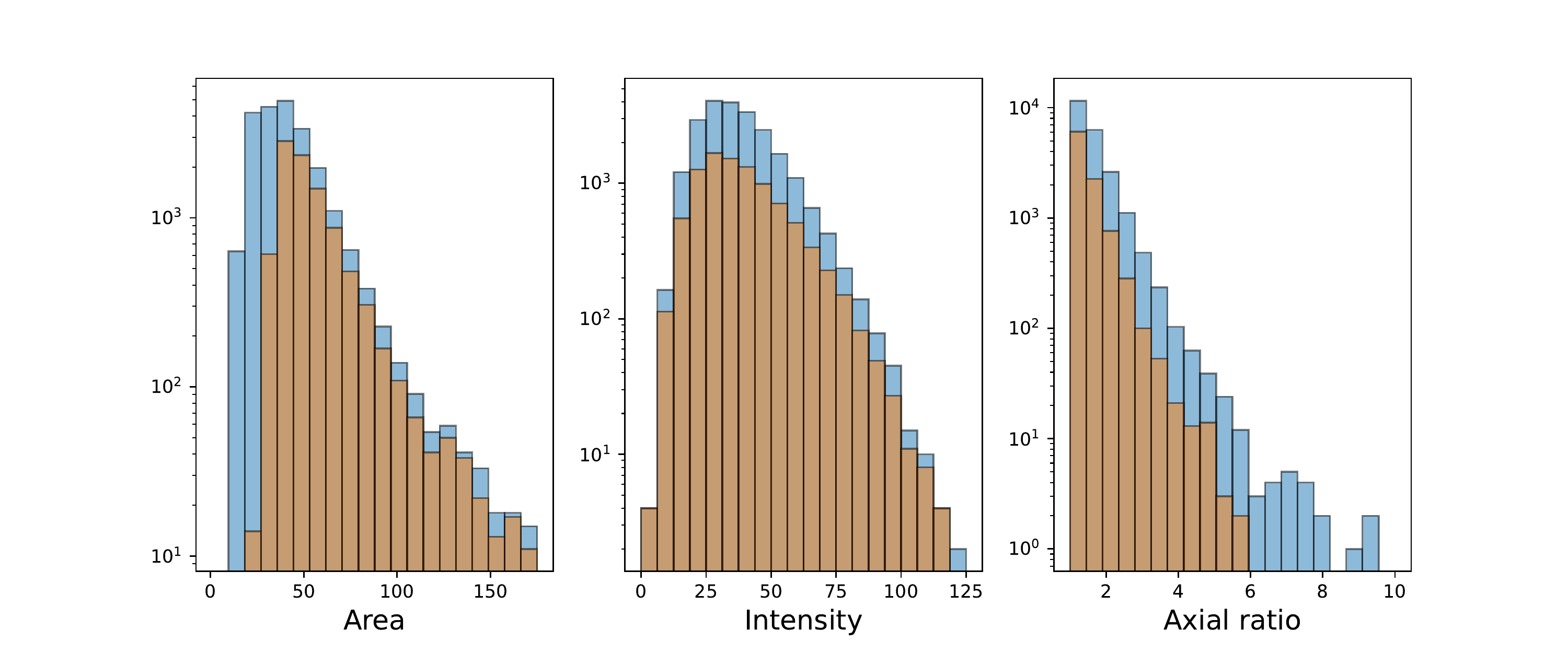} }}\\%
    {{\includegraphics[width=14.5cm]{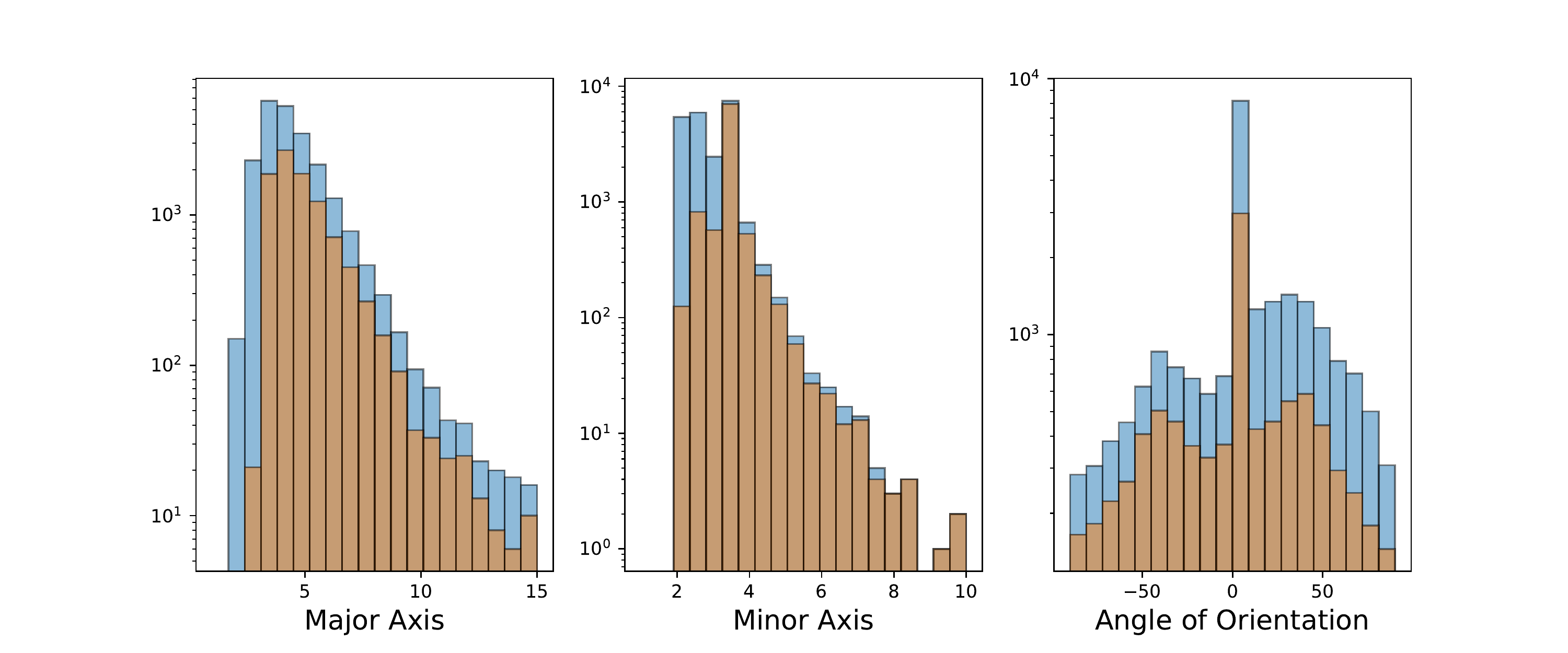} }}%
    \caption{Histograms of characteristics for Dataset 20200620 with frequency 120.52 MHz. Blue denotes all the peaks fitted, orange denotes the peaks that are dissimilar to the psf.}
    \label{fig:hist12-15}
\end{figure*}

A key objective of developing this robust technique for detection of WINQSEs was to go beyond the earlier works and characterise their spatial morphology in some detail.
For morphological characterisation, it is very important to deconvolve the detected WINQSEs from the restoring beam used during the imaging process. 
As both the detected WINQSEs and the restoring beam can be modelled well by Gaussians, the true size of the WINQSEs can be estimated by removing the effect of the restoring beam. For one dimensional Gaussians, the relation is straightforward:
\begin{equation}
    \sigma_{obs}^2 = \sigma_{restr}^2 + \sigma_{true}^2,
\end{equation}
where $\sigma_{obs}$, $\sigma_{restr}$ and $\sigma_{true}$ are the observed standard deviations of the Gaussian WINQSEs, the restoring beam and the true size of the WINQSEs respectively. This relation can be generalized for two dimensional Gaussian features with standard deviations $\sigma_x, \sigma_y$, oriented at an angle $\theta$ to the x-axis, and used to obtain the size of the true WINQSEs from the sizes of the observed features and restoring beams. For the observed features with $\sigma_x = \sigma_y, \ \theta = 0$, the relationship reduces to the simple additive relation above.   
{For sources much smaller than the resolution of the measurement, $\sigma_{true}$ approaches a $\delta$ function.
To separate the unresolved sources from others, we use a fiducial threhold of $2$ pixels$^2$.}
This translates approximately to a threshold of $\sim 40$ pixels$^2$  on the observed area, which is the size of the effective restoring beam. 
Figure \ref{fig:hist12-15} shows the parameters of the best fit Gaussian before deconvolution. 
The distributions for all of the WINQSEs which could be fit well by Gaussian models are shown in blue.
The peak of the distribution of area lies close to the area of the restoring beam, at $\sim 49$ pixel$^2$.
The distributions for the subset deemed to be resolved (or dissimilar to the restoring beam) are shown in orange. 
We note that while most of the WINQSEs with areas less than the beam size are considered unresolved, some with areas $ < 40$ pixel$^2$ are still considered resolved.
This is due to the complex non-linear relationship between the true and observed two-dimensional Gaussians. Similarly, we note that some features with large major axes ($> 10$ pixels) are considered unresolved. These features typically have minor axes much smaller than the beam size, hence, while they are not similar to the restoring beam, having one axis much smaller than the beam size results in their being regarded as an unresolved feature.

It is evident from this figure that only a small fraction of the detected WINQSEs are point source-like, the peak of the area distribution is very close to the area of the effective resolution, which is equal to $\sim 49$ pixel$^2$. 


\begin{figure*}
    \centering    
    {{\includegraphics[width=14.5cm]{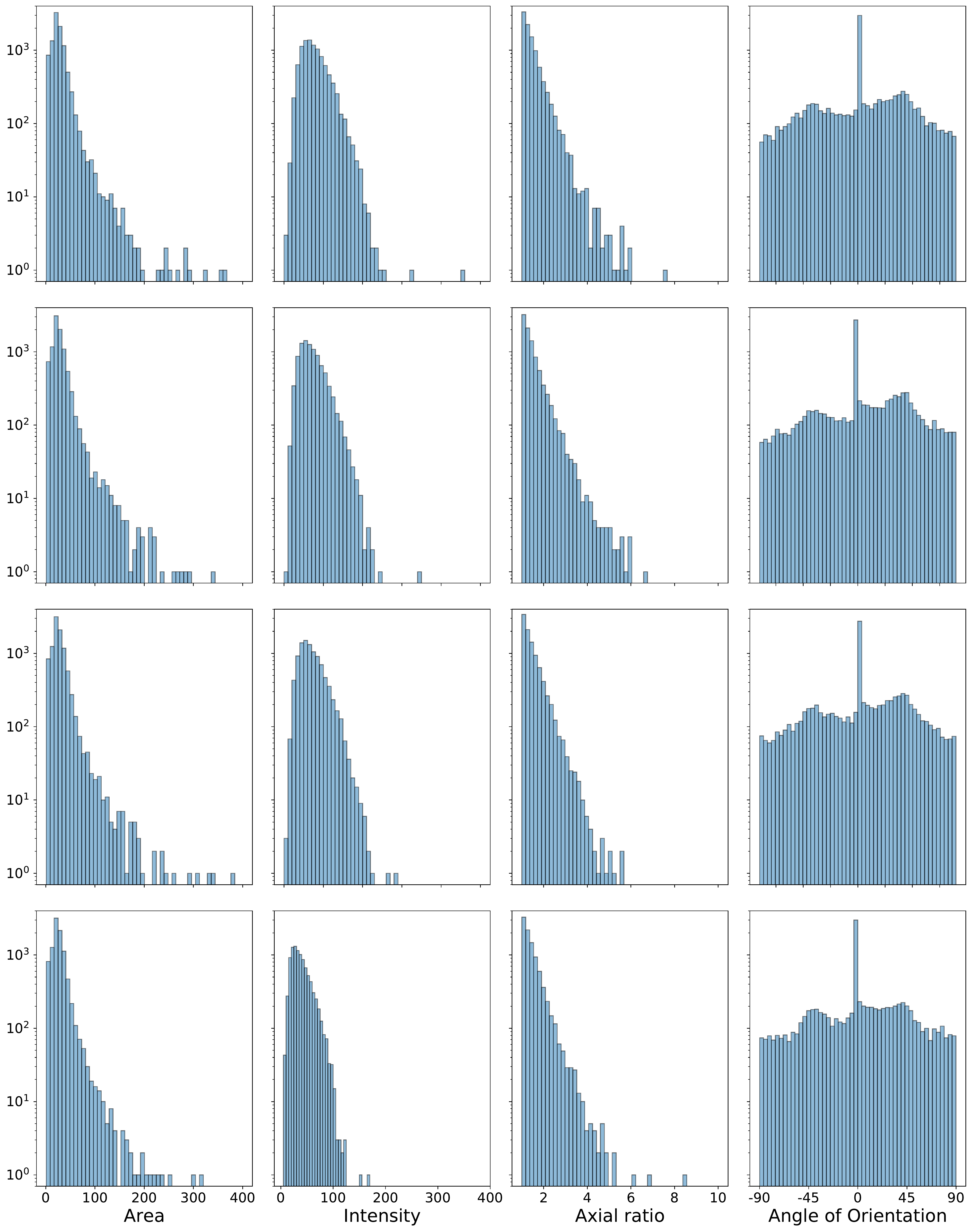} }}
    \caption{Histograms of best fit Gaussian parameters for Dataset 20200620 for 120.52 MHz, 128.20 MHz, 135.90 MHz, 143.60 MHz (top to bottom) for peaks dissimilar to the psf.}
    \label{fig:histQS}
\end{figure*}

\begin{figure*}
    \centering
    {{\includegraphics[width=13cm]{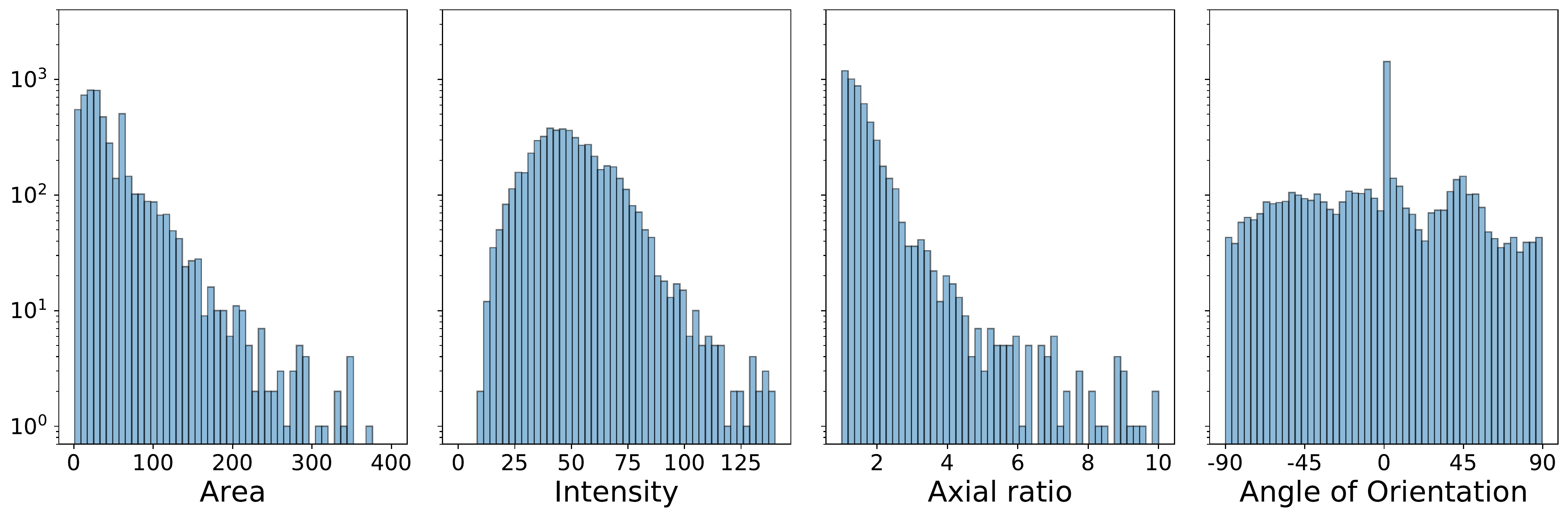} }}
    \caption{Histograms of characteristics for Dataset 20171127 with frequency 132 MHz for peaks dissimilar to the psf.}
    \label{fig:histAS}
\end{figure*}

Figures \ref{fig:histQS} and \ref{fig:histAS} show the distribution of various morphological parameters corresponding to the datasets 20200620 and 20171127 respectively after deconvolution. 
The first through fourth columns shows the distribution of area, intensity, axial ratio and the position angle of the deconvolved Gaussians. 
We find that the distributions are highly peaked in nature with the occurrence number often dropping by approximately two orders of magnitude within a factor of two in the value of the corresponding parameter value. 

We observe that in the four frequencies corresponding to 20200620 studied in this work, the area distribution shows a peak near the effective angular resolution of the data. 
We also note that the deconvolved area of the WINQSEs (Fig. \ref{fig:histQS}) has a distribution which is much more sharply peaked than that of the fitted area (Fig. \ref{fig:hist12-15}), hinting towards the fact that the WINQSEs are intrinsically compact in nature.
The area distribution in Fig. \ref{fig:histAS} on the contrary is much broader as is its peak. The range of distribution of the axial ratio is also large. 
We note that the Gaussians with high axial ratios are located preferentially close to the active region. 
While the distributions do contain a few outliers, on visual examination we find that they lie on the boundary of the various filters used and are pathological in nature.
They can hence be safely discarded.


%

\section{Discussion}
\label{sec:disc}

\begin{figure}
    \centering
    \includegraphics[scale=0.25]{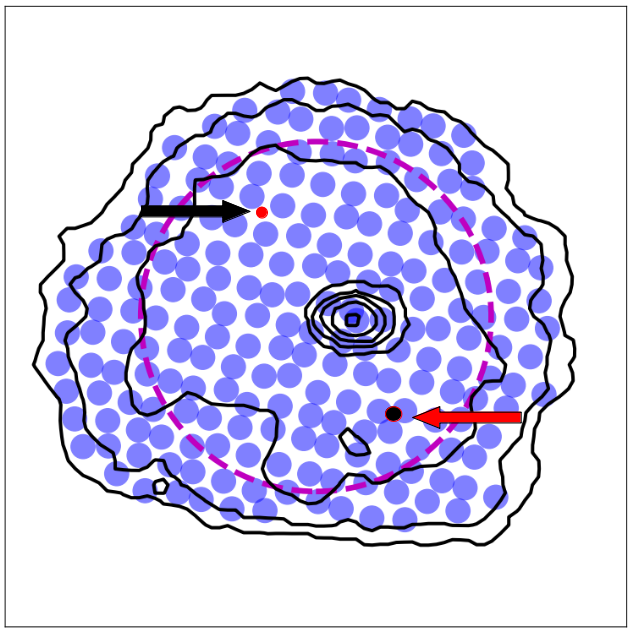}
    \caption{The purple circles mark the actual regions used by M20 for detecting WINQSEs. The red and black dots show two example location of WINQSEs chosen to demonstrate the weaknesses of the detection method employed by M20.}
    \label{fig:m20_regions}
\end{figure}

M20 discovered the presence of nonthermal radio transients from the quiet sun and showed that these emissions are ubiquitous in nature. 
They determined the flux density enclosed in several arbitrarily chosen psf sized regions on the quiet Sun and found that the distribution of the normalised median subtracted flux densities of these regions have a powerlaw tail, which is not expected from a thermal distribution. 
M23 confirmed the ubiquity of these features over the quiet sun in data taken during exceptionally quiet solar conditions, but found the distribution of median subtracted flux densities to be described by a log-normal distribution.
However, the technique employed both by M20 and M23 has some limitations, including the following.
It works with arbitrarily defined psf sized regions tiling the Sun, with no consideration to the image plane characteristics of these WINQSEs. This leads to two issues -- 1.) By regarding all features, even when present in adjacent psf sized regions, as independent, this approach intrinsically assumes that these emissions must be unresolved by these observations and hence compact in their morphology. 
2.) It is only to be expected that, even if these features were all to be unresolved, their locations would not align with centres of the psf sized regions chosen to tile the solar disc. This implies that a given feature could spill into multiple psf sized regions, leading, on the one hand, to an over-counting of events and, on the other, to underestimating their flux densities or perhaps not detecting them at all. 
These effects are illustrated in Fig. \ref{fig:m20_regions} using the regions employed by M20 and two appropriately chosen example locations for WINQSEs.
If a WINQSE is centered at the red dot (indicated using a black arrow), parts of its flux density will get distributed between its our neighbouring regions.
Given their low flux density, it is, hence, less likely to push the flux density at either these regions above the threshold for detection of WINQSEs and likely to remain undetected using the technique employed by M20. 
On the other hand, a somewhat stronger WINQSE located at the black dot (indicated using a red arrow), can potentially push the flux density of rhe three neighbouring regions above the threshold and lead to over-counting in the number of WINQSEs.  

Additionally, M20 did not use any threshold on the noise level in the image, but instead used a threshold for the dynamic range of the image, which is not as reliable an indicator of imaging quality as the noise. 
This approach leave open the rare possibility of spurious instances, when images meet the dynamic range criterion but also have noise high enough to contribute to the powerlaw tail.
    

By identifying these features in the image plane we address these issues and, for the first time, also extract morphological information about these features. We have implemented a machine learning algorithm which apart from filtering out noise and clustered WINQSEs, intelligently chooses various parameters so as to obtain the best possible Gaussian fit to the data. 

\subsection{The orientation of resolved WINQSEs}
\label{subsec:orientation}

Our initial hypothesis, based on M20, \citet{mondal2021} and M23, is that WINQSEs are the radio counterparts of nanoflares.
This implies that the intrinsic size of WINQSEs should be very small and they should appear as unresolved sources in our radio images.
Additionally, if these WINQSEs are indeed much weaker cousins of the usual type III and/or type I solar radio bursts, they must arise due to plasma emissions process at the local plasma frequency or its harmonic. 
It is also well established that emissions at low radio frequencies get modified significantly as they propagate through the turbulent and inhomogeneous coronal medium. 
The most prominent of these propagation effects is scattering.
Anisotropic scattering in the magnetized corona is kown to lead to an increase in the apparent angular size of the source, change the apparent morphologies of compact sources to appear non-circular and even shift the location of the centroid of emission from its true location \citep[e.g.][etc.]{arzner1999,mohan2019,kontar2019,murphy2021}.

Figures 16 and 17 show that the areas of the vast majority of the observed WINQSEs are larger than the rather generous threshold of 40 pixel$^2$ used for identifying resolved WINQSEs. 
We hypothesize that while WINQSEs are intrinsically compact and unresolved, their large apparent sizes with a significant fraction at axial ratios well beyond unity arise due to the anisotropic coronal scattering they suffer.
This then presents the possibility of using the observed morphologies of WINQSEs, especially their sizes and axial ratios to glean information about the coronal scattering.
In fact, an approach based on a similar idea, has been used successfully by \citet{Bastian1999, Kobelski2016}.
They used the Jansky VLA at 1--19 GHz to examine the apparent morphologies of compact extra-galactic radio sources as they passed close to the Sun (2--10 $R_{\odot}$, where $R_{\odot}$ is the solar radius). 
They find that practically all observed sources show angular broadening perpendicular to the radial direction.
At these heights the radial direction is also the expected direction of the magnetic field and the observed behaviour is consistent with expectation for kinetic Alfven waves.
 

\subsection{Comparison with earlier works}

M23 also analysed one of the same data presented here (2020620), though 
their technique for detecting WINQSEs was completely unrelated to the method followed here.
It is, hence, important to compare the statistical properties of the WINQSEs detected using the two methods to investigate the robustness of the results.

\begin{figure*}
    \centering
    \includegraphics[scale=0.5]{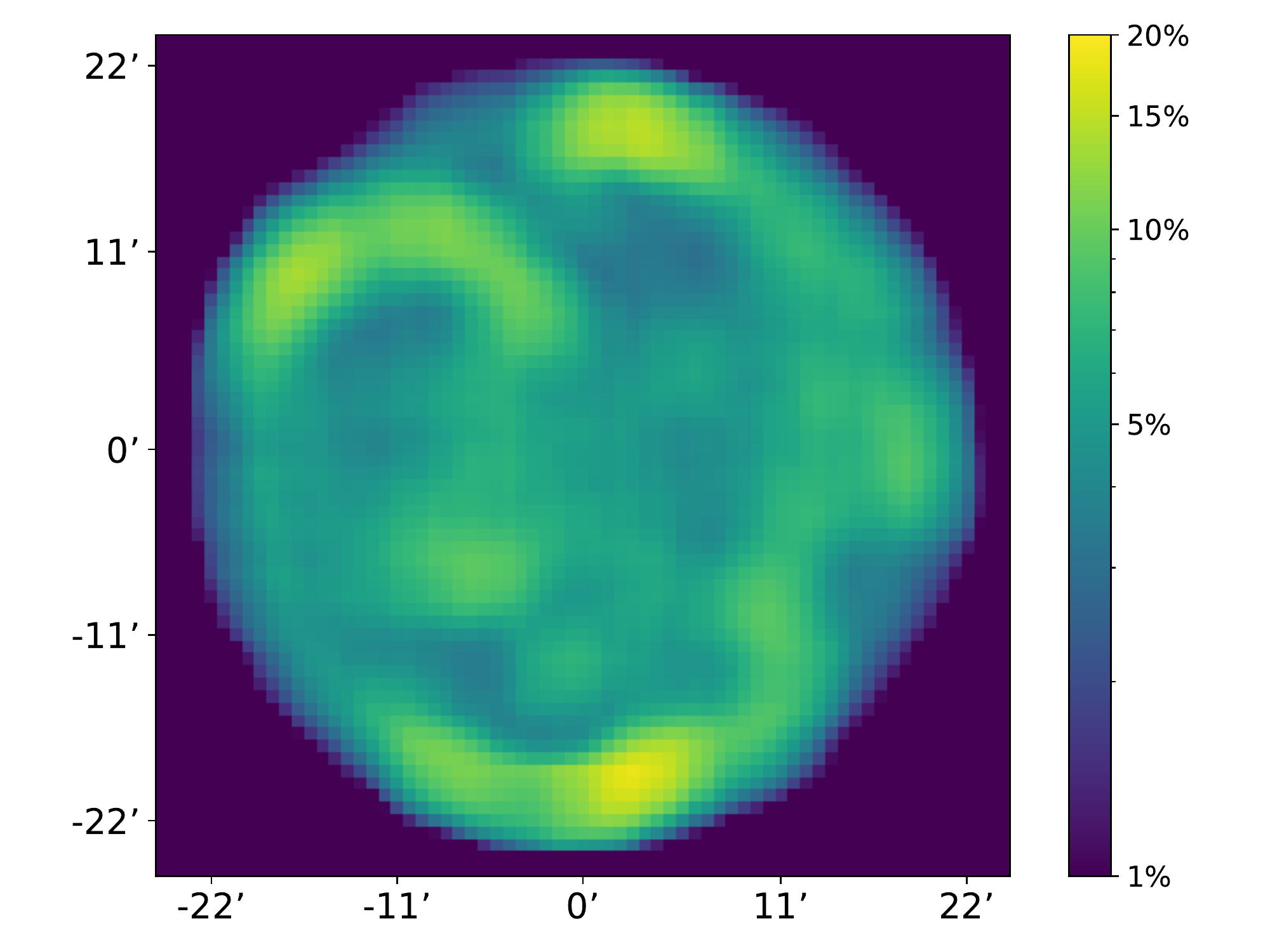}
    \caption{Shows the data presented in Fig. \ref{fig:incidence} at a resolution similar to that shown in M23}.
    \label{fig:ayan_occupancy}
\end{figure*}

From Fig. \ref{fig:incidence}, it is evident that the WINQSEs are detected everywhere on the sun. However, the fraction of time for which WINQSEs were detected on any given patch on the Sun is much lower than that presented in M23. 
This difference can easily arise from the simple fact that while M23 counted WINQSEs within psf sized patches, Fig. \ref{fig:incidence} counts the occurrence of WINQSEs over regions much smaller than the psf.
To demonstrate this quantitatively, Fig. \ref{fig:ayan_occupancy} plots the same occupancy distribution as shown in Fig. \ref{fig:incidence} after convolving it with a square kernel of size 7pix$\times$7pix, which is the approximate size of the regions used in M23. We observe that after this procedure, the values are consistent with those obtained by M23, thus validating our hypothesis regarding the reason behind the apparent discrepancy between Fig. \ref{fig:incidence} and that shown in M23.

Figure \ref{fig:deltaFbyF} shows the $\Delta F/F$ distribution of the WINQSEs detected here.  It is interesting to note that the in spite of very different analysis strategies the histogram of the intensities presented here is very similar to that reported in M23, demonstrating the robustness of the results presented here. 
There are, however, some small differences between the histograms presented here and in M23. For example, in this work, a few WINQSEs with $\Delta F/F \sim 10^{-2}$ 
have been detected. We believe that these WINQSEs were missed in M23, as it defined the psf sized regions within which to detect WINQSEs independent of the actual locations of individual WINQSEs.
Taking into account their true position, size and orientation when modeling them as Gaussians, enables us to detect weaker WINQSEs than were possible earlier. 

{\color{green}
}

\subsection{Importance of clustering and tolerance towards small errors}
\label{sec:clust}

The quality of the Gaussian fit depends on the choice of the fitting window used.
Determining a fitting window independently for each source is certainly possible, but leads to a larger computational burden. 
Also, as a robust fit should be insensitive to the small variations in the choice of fitting window, individually tuned windows for each feature should not be required either, and a fitting window close to the optimal should suffice. 
The approach presented here uses the DBSCAN algorithm on the dimensionally-reduced output of the t-SNE algorithm to determine an appropriate fitting window for the detected sources.
In view of the limitations of the DBSCAN algorithm in precisely identifying clustering, it is important to examine the final results of the Gaussian fitting process for their sensitivity to the details of the choice of the fitting window.
We note that the clustering properties are used only for the limited purpose of determining the fitting window, the rest of the procedure for fitting a 2D Gaussian is independent of this choice.
Here we examine the sensitivity of the final results from the Gaussian fitting to our approach of using DBSCAN on the dimensionally-reduced output of the t-SNE algorithm.

We fit all of the detected sources using the fitting window appropriate for Group 1 defined in Sec. \ref{subsec:optWindowSize} and define this group to be the reference cluster. 
We define a quantity $\Delta \chi^2=\chi^2_{ref}-\chi^2_{original}$, where $\chi^2_{original}$ refers to the $\chi^2$ obtained using the fitting window chosen based on the location of a source in the t-SNE plane as in Sec. \ref{subsec:optWindowSize} and $\chi^2_{ref}$ the results from the Gaussian fitting exercise using a fitting window corresponding to the reference cluster for all of the detected sources.
\begin{figure*}
\centering
\includegraphics[trim={0 0 12cm 0},clip,scale=0.55]{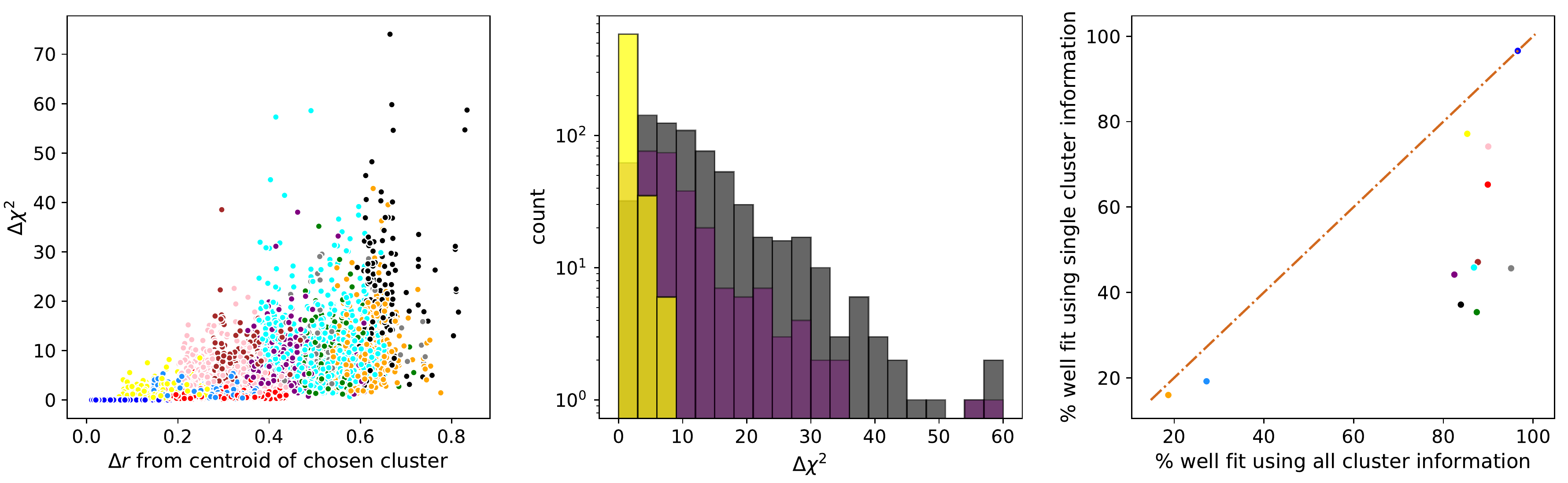}
\includegraphics[trim={24cm 0 0 0},clip,scale=0.55]{fits.pdf}
\caption{ Top left panel: $\Delta \chi^2$ is plotted against the distance of the point from the centroid of the reference cluster in the t-SNE plane. Top right panel: Histogram of $\Delta \chi^2$ for three chosen clusters, yellow (adjacent to the reference cluster), black (farthest cluster from the reference cluster) and purple (a cluster at an intermediate distance). Bottom panel: Shows the fraction of good fits. The X-axis corresponds to the fitting results shown in the manuscript where the fitting window is chosen based on the mean hyper-parameters of the cluster to which the point belongs. The Y-axis corresponds to the case when the same fitting window is used for all points. The orange line defines the case where the original fit and reference fit are identical. The color scheme for all panels is same as that in Fig. \ref{fig:groups}.}
\label{fig:fits}
\end{figure*}
The top left panel of Fig. \ref{fig:fits} shows the variation of $\Delta \chi^2$ with the distance between a given feature and the centroid of the reference cluster in the t-SNE plane, with the colors identifying the clusters as in Fig. \ref{fig:groups}.
The blue features are expectedly fit equally well ($\Delta \chi^2 = 0$), since they are fit using the same window as previously. For all other clusters, we note that the none of the features fit better with the new fitting window. The red, yellow, and light blue features, which come from clusters adjacent to the blue cluster, show a smaller degradation in fit quality while the features from faraway clusters fare much worse. While there are some features for which $\Delta \chi^2\approx 0$, in general the features from faraway clusters tend to lie at larger values of $\Delta \chi^2$. 
The top right panel shows the same results for three example clusters -- yellow (which lies adjacent the blue cluster), purple (which lies further away), and black (farthest from the blue cluster). 
Most of the features in the yellow cluster lie at low $\Delta \chi^2$ values. While a smaller number of features are seen at low $\Delta \chi^2$ values for the purple and black clusters as well, there is an increasing number of features lying at larger values of $\Delta \chi^2$. 
Note that no filtering on goodness of fit has been applied in the top panels of Fig. \ref{fig:fits}.
The bottom panel shows the percentage of fits that can be classified as good fits using the criteria laid out in Sec. \ref{subsec:fitGaussian} using the original fitting windows for the various clusters and that used for the reference cluster.
While the blue cluster has retained the goodness of fit from before as expected, all other clusters show a worsening in the number of the good fits, implying a reduction in the number of detected WINQSEs. 
For some clusters, the number of good fits were low to begin with and drop even lower (e.g. orange and light blue clusters).
For some other clusters, for which the number of good fits was $> 80\%$ with the original choice of fitting window drops to $< 50\%$ for the window corresponding to the reference cluster.

This demonstrates that, firstly, the choice of an optimal window is essential for a high quality fit. Fitting a single window to all features causes a clear degradation in fit quality. Thus the need to use clustering to obtain a classification of the features. At the same time, the quality of fits is robust to small variations in the exact choice of the window. Features closer to the reference cluster are fit nearly as well by the reference cluster window as by their initial window. Thus, even though the tSNE+DBSCAN algorithm may not be able to provide very accurate clustering, it is able to separate the features sufficiently well that the results arrived at here are not affected by the limitations in classifying the features accurately.


\section{Conclusion}
\label{sec:conclusion}
This work forms a part of a larger effort devoted to establishing the reliability of detection of WINQSEs and describing their observed characteristics.
The observed distributions of intensities of WINQSEs and their spatial distribution over the solar disc is found to be consistent with those reported in M20, \citet{rohit2022} and M23 using independent techniques.
For the first time, we investigate the morpohologies of WINQSEs and find them to be well described by 2D elliptical Gaussian models. 
A robust machine learning based pipeline has been designed and implemented to reliably identify and model the large number of WINQSEs (tens of thousands) observed.
After appropriate pre-processing of the images, the pipeline implements an unsupervised clustering approach using t-SNE to embed the information about the morphology of reliably detected features in a two dimensional space. DBSCAN is then used to group similar features followed by fitting 2D Gaussian models to obtain the best-fit parameters.
The results from this pipeline, the distributions of best fit Gaussian parameters -- area, intensity, axial ratio and angle of orientation -- for all of the WINQSEs regarded to be resolved are also presented.
The distribution of angle of orientation does not show any peak, implying that there is no preferred orientation for WINQSEs.
The distribution of all other quantities are very sharply peaked.

Naively, our interpretation of WINQSEs being radio counterparts of the nanoflares, hypothesiszed to explain coronal heating, suggests that their morphologies should be consistent with compact unresolved sources.
However, we find the vast majority of WINQSEs to be spatially resolved and regard this to be a consequence of scattering in the inhomogeneous coronal medium.
This opens the possibility that the studies of morphologies of WINQSEs can serve as a powerful probe for studying anisotropic scattering in the quiescent coronal regions. 
Their ubiquitous and ever-present nature makes them especially valuable, as one does not need to wait for favorable special circumstances or events.
Our attempts to look for patterns in orientations of WINQSEs as evidence for persistent inhomogenities in the coronal medium, expected due to the orientation of the local magnetic fields, were inconclusive.
Though we did not specifically investigate it here, most of the WINQSEs have been found to be unresolved in time at the 0.5 s time resolution of the data.
As the scattering process imposes a tell-tale time profile on intrinsically impulsive emissions \citep[e.g.][]{arzner1999, mohan2019}, it would be very interesting to examine the time profile of WINQSEs and we intend to pursue this.

We believe that WINQSEs science, which requires high SNR detection of weak short-lived narrow-band emissions, pushes the current generation of instrumentation close to their limits.
All of the evidence thus far is consistent with their being radio counterparts of nanoflares.
The tantalizing potential implications of these studies for coronal heating makes them a very exciting line of investigations with the upcoming more sensitive interferometers like the MWA Phase III and the SKA1-Low.
With the expected availability of SKA1-Mid precursor instruments like the MeerKAT and ASKAP for solar observations in the near term, and the SKA1-Mid itself closer to the end of this decade, it would be very interesting to look for WINQSEs like emissions at higher frequencies, and hence lower coronal heights.


\begin{acknowledgements}
This scientific work makes use of the Murchison Radio-astronomy Observatory (MRO), operated by the Commonwealth Scientific and Industrial Research Organisation (CSIRO).
We acknowledge the Wajarri Yamatji people as the traditional owners of the Observatory site. 
Support for the operation of the MWA is provided by the Australian Government's National Collaborative Research Infrastructure Strategy (NCRIS), under a contract to Curtin University administered by Astronomy Australia Limited. We acknowledge the Pawsey Supercomputing Centre, which is supported by the Western Australian and Australian Governments.
DO and AB acknowledge support of the Department of Atomic Energy, Government of India, under the project no. 12-R\&D-TFR-5.02-0700. SM acknowledges support by USA NSF grant AGS-1654382 to the New Jersey Institute of Technology.
The SDO is a National Aeronautics and Space Administration (NASA) spacecraft, and we acknowledge the AIA science team for providing open access to data and software. 
This research has also made use of NASA's Astrophysics Data System (ADS). 
We thank the developers of Python 2.7\footnote{See
\url{https://docs.python.org/2/index.html}} and Python 3\footnote{See \url{https://www.python.org/download/releases/3.0/}} and the various associated packages, especially Matplotlib\footnote{See \url{http://matplotlib.org/}}, Astropy,\footnote{See \url{http://docs.astropy.org/en/stable/}}, SciPy\footnote{See \url{https://www.scipy.org/}} and NumPy\footnote{See \url{https://numpy.org/}}. This research used version 4.0.5 \citep{mumford_2022_zenodo} of the SunPy open source software package \citep{sunpy_community2020}.
This research used version 0.6.4 \citep{barnes2020b} of the aiapy open source software package \citep{Barnes2020}.

\end{acknowledgements}

\begin{appendix}
\label{appendix}

\section{Sobel Edge Detection}
Sobel edge detection is an edge detection technique which performs a 2-D spatial gradient measurement on an image to highlight its edges, using Sobel filters. The Sobel filters are pair of $3 \times 3$ convolution kernels, designed to respond maximally to edges running vertically and horizontally to the grid:
\begin{eqnarray}
    G_x &=& 
\begin{bmatrix}
+1 && 0 && -1\\
+2 && 0 && -2\\
+1 && 0 && -1
\end{bmatrix} * I \nonumber\\
    G_y &=& 
\begin{bmatrix}
+1 && +2 & +1\\
0 && 0 & 0\\
-1 && -2 & -1
\end{bmatrix} * I \nonumber\\
    G &=& \sqrt{G_x^2 + G_y^2} \,\,,
\end{eqnarray}
where $I$ is the initial raw image and $G$ is the final image with edges highlighted. The kernels are applied separately to the input image, to produce separate measurements of the gradient component in each orientation, $G_x, G_y$, which are then combined to find $G$, the absolute magnitude of the gradient at each point. The large convolution kernel of the Sobel  smooths the input image and so makes the operator less sensitive to noise, leading to accurate gradient measurements. 

\section{t-SNE}
t-Distributed Stochastic Neighbor Embedding (t-SNE) is an unsupervised, non-linear technique for feature reduction. In this algorithm, one first measures the probability $p_{j|i}$ that the data points in higher dimensional space, ${\bf x}_i, {\bf x}_j$ are neighbours:
\begin{equation}
    p_{j|i} = \frac{exp(-||{\bf x}_i - {\bf x}_j||^2)/2\sigma^2}{\sum_{k \neq i} exp(-||{\bf x}_i - {\bf x}_k||^2)/2\sigma^2}\,\,,
\end{equation}
where the scaled squared Euclidean distance (``affinity'')  is used to calculate the distance between the points, and $\sigma$ is the bandwidth of the Gaussian kernels. A random dataset $Y= \{ {\bf y}_0, {\bf y}_1, {\bf y}_2, {\bf y}_3, .... {\bf y}_N \}$ is generated in lower dimension (usually two or three dimensions), and the probability $q_{ij}$ of data points in $Y$ being neighbours to each other is given by
\begin{equation}
    q_{ij} = \frac{(1 + ||{\bf y}_i - {\bf y}_j||^2)^{-1}}{\sum_{k \neq l} (1 + ||{\bf y}_k - {\bf y}_l||^2)^{-1}} \,\,,
\end{equation}
where the probability is calculated using a Student t-distribution. The Student t-distribution is preferred over the normal distribution since its heavier tails allow for better modelling of far apart distances. The set of probabilities $q_{ij}$ is expected to be a good representation of the higher-dimensional probabilities $p_{j|i}$. To achieve this, the difference between the probability distributions is calculated using Kullback-Leibler divergence, and is reduced by minimizing the Kullback-Leibler cost function:
\begin{equation}
    C = \sum_{i}\sum_{j} p_{ij}log\frac{p_{ij}}{q_{ij}} \,\,.
\end{equation}
This ensures that the lower dimensional space is an adequate depiction of the initial higher dimensional feature space.

\section{DBSCAN}
Density-Based Spatial Clustering of Applications with Noise (DBSCAN) is an unsupervised machine learning algorithm \citep{ester1996density}. The strength of this algorithm is that it labels the outliers in the dataset, and can identify the number of clusters required based on the density of datapoints in a region.

Since this is a density based clustering method we need to provide two parameters:
\begin{enumerate}
    \item $\epsilon$ - The radius in which data points can be marked as neighbours of each other
    \item $MinPts=n$ - minimum number of neighbours  within the radius $\epsilon$ for a data point to be considered a core point.
\end{enumerate}

While clustering each data point is marked as either a core point, a boundary point or an outlier.
\begin{enumerate}
    \item core point: A data point which has neighbouring points  $\geq n$ within the radius of $\epsilon$.
    \item boundary point: A data point  which has neighbouring points  $< n$ within the radius of $\epsilon$, but is a neighbour of a core point.
    \item outlier: A data point which is not a neighbour of a core point, and has $< n$ neighbouring points. 
\end{enumerate}

The algorithm begins by first finding a core point within the dataset, and then extends its cluster from there on recursively. The points that are not core points or are not the neighbours of the core point are marked as noise. The extension of cluster happens by checking if any of the neighbouring points of the core point are themselves a core point as well. If they are then its neighbours would be in the same cluster and recursively it keeps finding such core points till the time it is only left with boundary points.



\end{appendix}

%
%

\software{CASA\citep{casa}, astropy\citep{astropy:2013, astropy:2018, astropy2022},  matplotlib\citep{Hunter:2007},  Numpy\citep{harris2020array}}, SciPy\citep{2020SciPy-NMeth}, pandas\citep{reback2020pandas}, SunPy \citep{sunpy_community2020}, AiaPy \citep{Barnes2020}

\facilities{Murchison Widefield Array \citep[MWA,][]{lonsdale2009,tingay2013,wayth2018}, Solar Dynamics Observatory \citep[AIA,][]{pesnell2012} }

\bibliography{references}   

\end{document}